\documentclass[twocolumn,10pt]{article}

\usepackage[utf8]{inputenc}
\usepackage[T1]{fontenc}
\usepackage{amsmath,amsfonts,amssymb,amsthm}
\usepackage{graphicx}
\usepackage{xcolor}
\usepackage{hyperref}
\usepackage{cite}
\usepackage{authblk}
\usepackage{geometry}
\usepackage{booktabs}
\usepackage{multirow}
\usepackage{tabularx}
\usepackage{algorithm}
\usepackage{algpseudocode}
\usepackage{tikz}
\usetikzlibrary{shapes,arrows,positioning,calc,patterns,decorations.pathreplacing}
\usepackage{booktabs}
\usepackage{xcolor}
\usepackage{colortbl}
\usepackage{pifont}
\usepackage{array}
\usepackage{ragged2e}
\usetikzlibrary{
  positioning,
  arrows.meta,
  fit,
  backgrounds
}

\usepackage{fancyvrb}

\definecolor{tlsblue}{HTML}{1A4D8F}
\definecolor{lightgrayrow}{gray}{0.95}

\newcommand{\cmark}{\textcolor{tlsblue}{\ding{51}}}
\newcommand{\xmark}{\textcolor{red!70!black}{\ding{55}}}

\hypersetup{
    colorlinks=true,
    linkcolor=blue,
    citecolor=blue,
    urlcolor=cyan,
}

\title{\textbf{Post-Quantum Identity-Based TLS for 5G Service-Based Architecture and Cloud-Native Infrastructure}}

\author[1]{Vipin Kumar Rathi}
\author[2]{Lakshya Chopra}
\author[1]{Nikhil Kumar Rajput}

\affil[1]{Ramanujan College, University of Delhi, New Delhi, India}
\affil[2]{coRAN Labs Private Limited, New Delhi, India}

\date{}

\begin{document}

\twocolumn[
\begin{@twocolumnfalse}
\maketitle

\begin{abstract}
Cloud-native application platforms and latency-sensitive systems such as 5G
Core networks rely heavily on certificate-based Public Key Infrastructure
(PKI) and mutual TLS to secure service-to-service communication. While
effective, this model introduces significant operational and performance
overhead, which is further amplified in the post-quantum setting due to large
certificates and expensive signature verification.

In this paper, we present a certificate-free authentication framework for
private distributed systems based on post-quantum Identity-Based Encryption
(IBE). Our design replaces certificate and signature based authentication
with identity-derived keys and identity-based key encapsulation, enabling
mutually authenticated TLS connections without certificate transmission or
validation. We describe an IBE-based replacement for private PKI, including
identity lifecycle management, and show how it can be instantiated using a
threshold Private Key Generator (T-PKG).

We apply this framework to cloud-native application deployments and
latency-sensitive 5G Core networks. In particular, we demonstrate how
identity-based TLS integrates with the 5G Service-Based Architecture while
preserving security semantics and 3GPP requirements, and we show how the same
architecture can replace private PKI in Kubernetes, including its control
plane, without disrupting existing trust domains or deployment models.
\end{abstract}

\end{@twocolumnfalse}
]

\section{Introduction}

This paper investigates whether certificate-based Public Key Infrastructure
(Private-PKI) remains a necessary component of authentication in cloud-native,
private distributed systems in the post-quantum era. We argue that it does not.

Modern cloud-native applications and 5G Core networks operate in centrally
administered environments where identities are already defined, scoped, and
authorized by the system itself. In the 5G Service-Based Architecture (SBA),
Network Functions are realized as microservices that communicate via
authenticated HTTP/2 APIs, often invoking multiple short-lived,
latency-sensitive connections per procedure. Despite this, authentication in
these systems relies on private PKI and mutual TLS, requiring certificate
issuance, validation, and lifecycle management for each participating service.

The transition to post-quantum cryptography fundamentally increases the cost
of this approach. Post-quantum signature schemes significantly inflate
certificate sizes and verification overhead, while certificate chains and
revocation mechanisms impose additional bandwidth and latency costs. In
systems such as 5G Core networks—where authentication is performed repeatedly
and latency budgets are tight—these costs directly impact operational
efficiency.

Crucially, this overhead does not arise from changing security semantics.
Authentication in these environments remains the verification of possession
of a system-recognized identity. Certificate-based PKI re-encodes identity
information the system already maintains, introducing indirection and
operational complexity that is structurally unnecessary in private domains.

We propose replacing private PKI with identity-based authentication using
post-quantum Identity-Based Encryption as the cryptographic foundation. In an
identity-based system, public keys are derived directly from identity strings,
eliminating the need for certificates altogether. Authentication is achieved
implicitly through identity-based key encapsulation: only an entity holding
the correct identity private key can successfully establish a secure channel.

We instantiate this approach as post-quantum identity-based TLS (IBE-TLS),
including a mutually authenticated variant suitable for systems that
currently rely on mTLS. The protocol preserves the TLS~1.3 record layer and
key schedule while removing certificate issuance, distribution, and signature
verification from the authentication path.

We apply this model first to cloud-native application workloads and
latency-sensitive 5G Core networks, demonstrating how identity-based TLS
integrates naturally with the Service-Based Architecture while preserving
3GPP security requirements and communication semantics. We then show how the
same architecture applies to Kubernetes itself, replacing its private PKI for
control-plane and component authentication while preserving existing trust
domains, bootstrapping semantics, and authorization models.

\paragraph{Contributions}
This paper makes the following contributions:
\begin{itemize}
  \item We present a certificate-free authentication architecture for private
  distributed systems, based on post-quantum identity-based cryptographic
  primitives. This architecture replaces conventional Certificate Authorities
  with identity-based components such as a Private Key Generator within a
  PKI-like framework (IBE-based PKI).

  \item We design a \textit{post-quantum} identity-based TLS protocol (IBE-TLS),
  including a mutually authenticated variant that replaces certificate and
  signature based authentication. The protocol derives its post-quantum
  security from a lattice-based identity primitive obtained via a slight
  adaptation of ML-KEM (FIPS~203), while using unmodified ML-KEM for the TLS
  ephemeral key exchange.

  \item We apply the IBE-based PKI and IBE-TLS model to cloud-native,
  Kubernetes-based application workloads, focusing on the 5G Service-Based
  Architecture. We demonstrate certificate-free mutual authentication between
  Network Functions while preserving 3GPP security requirements.

  \item We further show that the same approach applies to Kubernetes itself,
  where the private PKI used by control-plane components is replaced by an
  identity-based architecture that integrates with existing trust domains,
  identities, and lifecycle mechanisms. This framework is used to demonstrate
  mutually authenticated IBE-TLS between Kubernetes components while
  preserving the same security guarantees.
\end{itemize}

The remainder of this paper develops this architecture in detail, beginning with the design of IBE-TLS and the associated identity-based PKI, and its application to cloud-native deployments and 5G Core networks, culminating in an IBE-based replacement for the Kubernetes control-plane PKI.

\section{Motivation}

\subsection{Authentication in Private Distributed Systems}

Many modern distributed systems operate in private, centrally
administered environments. Enterprise microservice platforms, cloud
infrastructure, telecom core networks, and edge deployments are
typically owned and managed by a single organization. In these systems,
identities are explicitly defined, scoped, and enforced by the platform
itself. 
Components do not discover trust dynamically; instead, trust is
established through out-of-the-band configurations, authorization and access policies, and component names.

Despite this, authentication in such systems is almost universally
implemented using Public Key Infrastructure (PKI). Secure communication
is achieved by issuing X.509 certificates to components and validating
certificate chains during protocol handshakes. This approach inherits
assumptions from the public Internet setting, where identities are
external, long-lived, and must be verified across administrative
boundaries.

In private systems, however, these assumptions do not hold. Identities
are already known to the system, and the role of authentication is not
to establish who a peer claims to be, but to verify that it is the
authorized holder of a system-defined identity.

\subsection{Private PKI as a Mismatch}

Using PKI in private deployments introduces an architectural mismatch.
Certificates re-encode identities that are already managed by the
system, while certificate validation re-checks properties that are
already enforced through access control and configuration. As a result,
certificates function as an intermediate artifact rather than a primary
security mechanism.

This mismatch manifests operationally. Certificates must be issued,
rotated, stored, distributed, and revoked, often requiring dedicated
infrastructure and automation. Trust anchors must be consistently
propagated, and failures in certificate management frequently result in
service disruption.

\subsection{Authentication in 5G Core Networks}
The 5G Core Network (CN) is a representative example of a latency-sensitive,
cloud-native distributed system. It adopts a Service-Based Architecture (SBA)
in which Network Functions (NFs) are implemented as microservices that
communicate via RESTful APIs over HTTP/2 and TLS. Core procedures such as UE
registration, session establishment, mobility handling, and service discovery
involve multiple sequential NF-to-NF interactions, each requiring mutual
authentication.

These interactions occur on every subscriber procedure and are therefore
performance-critical. While classical mutual TLS authentication incurs
moderate overhead, the transition to post-quantum cryptography fundamentally
changes this cost structure. Post-quantum signature schemes significantly
increase authentication data size and verification cost, making certificate-
based mutual TLS a dominant contributor to latency and bandwidth overhead in
SBA procedures.

Importantly, this overhead does not arise from changing security semantics.
All Network Functions belong to a single administrative domain, their
identities are centrally defined and managed, and trust relationships are
established by configuration rather than discovery.

\subsection{Post-Quantum Cryptography Exposes the Problem}

The transition to post-quantum cryptography amplifies this mismatch.
Post-quantum signature schemes significantly increase certificate sizes
and verification costs, inflating TLS handshakes by several kilobytes
per authenticated peer and introducing additional computational
overhead. In mutual authentication settings, these costs are incurred by
both parties.

While such overhead may be tolerable in long-lived or data-intensive
connections, it becomes problematic in systems characterized by
frequent, short-lived interactions. In these environments,
authentication overhead can exceed application payload sizes and
directly impact latency-sensitive control-plane operations.

Post-quantum migration therefore forces a reconsideration of whether
certificate-based authentication remains appropriate for private
systems, rather than simply replacing classical primitives with larger
post-quantum equivalents.

Table \ref{tab:sign_sizes} lists the authentication overhead because of signature schemes, providing a stark comparison between classical and post-quantum signature schemes.

\begin{table*}[h!]
\centering
\small
\renewcommand{\arraystretch}{1.2}
\hspace{-1cm}
\begin{tabular}{|l|>{\columncolor{lightgray}}c|}
\hline
\textbf{Scheme} & \textbf{Key + Signature Size (KB)} \\
\hline
ECDSA                   & 0.1 \\
RSA                     & 0.5 \\
Lattice-based            & 11 \\
Stateful HBS            & 15 \\
Stateless HBS           & 42 \\
ZK Proofs (ex: Picnic L1FS) & 66\\
Multivariate            & 100 \\
Supersingular Isogenies & 122 \\
Code-based              & 190 \\
\hline
\end{tabular}
\caption{Estimates for key and signature sizes digital signature schemes.}
\label{tab:sign_sizes}
\end{table*}

\subsection{Identity-Based Authentication as an Alternative}

Identity-based cryptographic systems provide a structurally different
approach. Public keys are derived directly from identities, and
authentication is achieved by proving possession of an identity-bound
private key. Certificates and signatures are no longer required, and trust anchors are
reduced to a small set of system-wide public parameters.

In private deployments, where identifiers such as service names,
component roles, etc, are already defined and
centrally managed, identity-based authentication aligns naturally with
system semantics. Secure channels can be established using identities
alone, without transmitting or validating certificate chains. This is an idea explored with the beginnings of IBE.

Replacing private PKI with identity-based authentication collapses
several layers of infrastructure into a single abstraction: \textbf{identity}.
Certificate issuance becomes identity key extraction, certificate
validation becomes implicit authentication via key possession, and
revocation becomes a function of identity lifecycle management rather
than external revocation mechanisms.

\subsubsection{No pre-distribution of public keys}
A key practical property of identity-based systems is that both public
and private keys can be generated on demand, rather than pre-generated
and distributed ahead of time ~\cite{rfc5408}. A sender can derive a recipient's public
key directly from its identity and the public parameters of the system, without
any prior interaction, directory lookup, or certificate exchange. This
is in contrast to conventional public-key systems, where public keys
must be generated in advance, bound into certificates, and distributed
before secure communication can begin. In private systems, this
\emph{just-in-time} key derivation model aligns naturally with dynamic
workloads and ephemeral identities, removing the need for advance key
distribution or public-key directories.
\footnote{However, we do note that the pre-distribution of keys is not always required if using CA-based trust anchors.}

\subsection{Post-Quantum Identity-Based TLS}

The authentication guarantees of TLS, including mutual authentication,
can be realized using identity-based cryptography through identity-based
key encapsulation. By encapsulating a shared secret to an
identity-derived public key, a party implicitly authenticates the peer:
only the holder of the corresponding identity private key can
successfully decapsulate the secret.

When instantiated using a post-quantum identity-based KEM (ID-ML-KEM) \cite{id-ml-kem},
this approach removes certificates and digital signatures entirely from
the TLS handshake. Authentication is integrated directly into key
establishment, while the TLS record layer and key schedule remain
unchanged. Mutual authentication is achieved by performing identity-based encapsulation in both directions.

This motivates the design of a certificate-free, post-quantum
authentication architecture for private distributed systems.

\paragraph{Goals of this work.}
This work has the following concrete goals:
\begin{itemize}
  \item \textbf{Replace private PKI in centrally administered systems:}
  Eliminate the use of X.509 certificates, certificate chains, and
  certificate-based trust anchors in private deployments where
  identities are already well-defined and managed.

  \item \textbf{Adopt identity-based cryptography as the authentication
  primitive:}
  Use Identity-Based Encryption (IBE) to derive public keys directly
  from system-defined identities, removing the need for certificate
  issuance, distribution, validation, and revocation.

  \item \textbf{Design a post-quantum identity-based TLS protocol:}
  Instantiate authentication using a post-quantum identity-based key
  encapsulation mechanism (ID-ML-KEM), producing a certificate-free
  TLS handshake (IBE-TLS), which further preserves the existing TLS key schedule, record layer, and does not alter functionality.

  \item \textbf{Support mutual authentication without certificates:}
  Provide a mutually authenticated variant of IBE-TLS that serves
  as a direct replacement for certificate-based mTLS, achieving
  authentication implicitly through key encapsulation and
  decapsulation.

 \item \textbf{Transform 5G Core Network PKI and mTLS using IBE}: Use IBE-based PKI and IBE-TLS to serve as a replacement for private PKI systems used in 5G CN deployments, binding authenticity directly to the identities of individual components.

  \item \textbf{Demonstrate further applicability through Kubernetes:}
  Apply the proposed architecture to Kubernetes, replacing its private
  PKI with an identity-based system while preserving trust domains,
  component identities, and lifecycle semantics.

  \item \textbf{Target private, not public, deployment environments:}
  Focus on systems operated under a single administrative domain,
  where identity issuance via a (threshold) PKG is an acceptable and
  realistic trust assumption.
\end{itemize}

\section{Related Work}

\subsection{KEM-Based Authentication in TLS}

Several recent proposals replace signature-based authentication in TLS
with KEMs. The AuthKEM draft by Celi, Schwabe,
and Stebila~\cite{schwabe2023authkem} introduces a KEM-based TLS
authentication framework in which authentication is achieved through
successful decapsulation rather than digital signatures. These designs
reduce reliance on signatures but continue to assume long-term public
keys distributed via certificates or equivalent mechanisms.

These approaches reduce reliance on signatures but continue to assume
certificate-based public key distribution and trust anchors. In contrast,
our work eliminates certificates entirely by deriving public keys directly
from identities.

\subsection{Identity-Based Encryption}

Identity-Based Encryption was introduced by Boneh and Franklin using
bilinear pairings~\cite{boneh2001identity}. To achieve post-quantum
security, lattice-based IBE constructions were proposed by Gentry,
Peikert, and Vaikuntanathan using trapdoor sampling techniques
\cite{gentry2008trapdoors}, with subsequent extensions to hierarchical
and more expressive settings~\cite{agrawal2010lattice}.

These works establish the cryptographic foundations of IBE but do not
consider its integration into transport-layer security protocols or
large-scale distributed systems. Our work builds on lattice-based IBE
and adapts it specifically for TLS authentication and cloud-native
deployment.

\subsubsection{Identity-Based TLS}

Identity-based authentication mechanisms for TLS have been explored in
both classical and post-quantum settings. Prior work has shown that IBE
can replace certificate-based authentication by embedding identity-based
key agreement into TLS handshakes. These proposals demonstrate the
feasibility of certificate-free authentication at the protocol level. Post-Quantum instantiations of IBE-TLS using Lattice-based primitives, has been studied for enhancing efficiency and reduce overheads in Post-Quantum TLS mechanisms \cite{banerjee2020efficient}. Classical constructions didn't offer any sufficient overhead reduction considering the classical signatures themselves --- RSA, ECDSA, Ed25519, had low signature and key sizes.

However, existing IBE-TLS designs primarily focus on cryptographic
protocol construction and do not address system-level deployment,
identity lifecycle management, or integration with modern orchestration
platforms. In particular, they do not consider how identity issuance,
rotation, and trust domain separation interact with real-world systems
such as Kubernetes.

\subsection{Threshold Private Key Generator and Lattice Trapdoors}

Key escrow is a known limitation of IBE systems. Threshold constructions
mitigate this risk by distributing the master secret key across multiple
authorities. Threshold lattice trapdoor generation and distributed key
extraction have been studied extensively, such as ~\cite{boneh2018threshold, BendlinKrehbielPeikert2013}.

We build on these techniques to present a Threshold Private Key Generator (T-PKG) architecture for cloud-native systems, which is fully compatible with IBE-TLS, and allows for greater trust and improved fault tolerance.

\subsection{Authentication and PKI in Cloud-Native Systems}

Cloud-native platforms such as Kubernetes and 5G Core deployments rely on
private, certificate-based Public Key Infrastructure (PKI) to secure
communication between control-plane components and application workloads.
Prior work has primarily focused on improving PKI operations, including
automated certificate issuance and rotation, hierarchical trust structures,
and ACME-style provisioning mechanisms.

In contrast, this work does not optimize certificate-based PKI, but replaces
it entirely using identity-based authentication. To our knowledge, no prior
work has proposed eliminating private PKI in cloud-native systems—including
Kubernetes and the 5G Service-Based Architecture—while preserving existing
identity, authorization, and lifecycle semantics.

\subsection{Summary}

In summary, while KEM-based TLS authentication, identity-based
cryptography, and Kubernetes PKI have each been studied independently,
this work is the first to combine post-quantum identity-based TLS with a
cloud-native system architecture, such as 5G Core and Kubernetes, addressing certificate
management overhead at both the cryptographic and systems levels.

\section{Background}

\subsection{Identity-Based Encryption}
\subsubsection{IBE cryptosystem and PKI}
Identity-based encryption is a public-key cryptosystem where the encryption procedure is carried out using public keys derived from user identities, such as their name, e-mail id, etc. In contrast, conventional public-key cryptosytems rely on authenticated distribution of public keys via the use of PKI components such as certificate authorities and digital signatures. In TLS 1.3 for instance, the server proves its authenticity using the end-entity certificate issued to it by the an intermediate or root CA. Authentication involves verification of the certificate chain from the end-entity to the root CA with additional signature verification over the entire TLS handshake transcript. IBE eliminates the need for such an infrastructure --- certificate authorities, certificate storage and verification. 
\subsubsection{Private Key Generator and Key Escrow}
In IBE, a trusted third-party called the Private Key Generator (PKG) holds a master secret key and can derive secret keys for arbitrary identities represented as strings. A user (or an end-entity) receives their secret key after authenticating themselves with the PKG in the same manner that they would do to a Certificate Authority (CA). This eliminates the need for a certificate-based public-key infrastructure. However, the generation of the private key by the PKG also leads to \textbf{key escrow}, since it knows private key of the end-entity. This is the primary limitation of IBE, since the PKG can derive secret keys for any identity and thus decrypt all communications. Threshold cryptography and Distributed Key Generation techniques provides a mitigation by distributing the master secret key across multiple servers such that no single server can extract identity keys alone.
\subsubsection{Classical constructions of IBE}
The problem of IBE cryptosystems despite being first proposed in 1984 had no practical solutions until Cocks IBE scheme and Boneh-Franklin's pairing-based scheme. Cocks built IBE scheme with security basis on the quadratic residuosity problem, which is believed to be intractable. Boneh-Franklin scheme \cite{boneh2001identity} on the other hand uses bilinear pairings on certain elliptic curve subgroups, which allows for a secondary private key (in addition to the master private key) which works only for a specific identity. The scheme is secure as long as the Computational Diffie-Hellman Problem is hard in the chosen group. Boneh-Franklin scheme is indistinguishable under adaptive chosen ciphertext attack for which a new security notion --- \texttt{IND-ID-CCA} is devised. ~\cite{DottlingGarg2017} introduces the first construction of IBE based only on the security of the Computational Diffie-Hellman problem or factoring, without relying on bilinear pairings. This construction uses garbled circuits for a non-black-box use of cryptographic primitives consequently bypassing the impossibility of construction of an IBE scheme using standard trapdoor permutations in a black-box manner \cite{boneh2008impossibility}. 


\subsubsection{Lattice-based constructions of IBE and HIBE}
Lattice-based cryptographic schemes achieve post-quantum security and offer low asymptotic complexity, making them attractive for a wide range of cryptosystems, including IBE. The works of Gentry, Peikert and Vaikunanthan \cite{gentry2008trapdoors} uses a master public key containing a matrix $\mathbf{A}$ and a master secret key containing a trapdoor for $\mathbf{A}$. To extract a key for identity $ID$, the PKG computes a matrix $\mathbf{A}_{ID}$ by hashing the identity and uses the trapdoor to sample a short vector $\mathbf{s}_{ID}$ such that $\mathbf{A}_{ID} \mathbf{s}_{ID} = \mathbf{u}$ for some target $\mathbf{u}$. This is structurally analagous to Cocks IBE scheme (quadratic residuosity) since they both employ a "pre-image sampleable" trapdoor function. Due to their reliance on the random oracle model for security analysis, they have faced both theoretical and practical drawbacks. 
An alternative lattice-based construction, known as the \textbf{Bonsai Trees} scheme \cite{CashHofheinzKiltzPeikert2009}, employs a hierarchy of trapdoor functions indexed by public lattices randomly sampled from a family of lattices whose hardness reduces to the Learning With Errors (LWE) problem; i.e., each identity corresponds to a lattice $\Lambda_{id}$ equipped with a short basis (trapdoor) that enables efficient sampling of preimages, while remaining computationally hidden from any adversary without the trapdoor. This hierarchy is used to realize a HIBE scheme that supports delegation across arbitrary depth while maintaining anonymity and security under standard lattice assumptions.
. The Bonsai Tree structure mirrors the certification hierarchy: the root authority begins with a master trapdoor for the top-level lattice and can derive child trapdoors by embedding identities as\textbf{ controlled modifications} of this lattice. Each delegated authority receives a short basis for its branch, enabling it to compute trapdoors for its descendants.

\subsection{Post-Quantum TLS 1.3}

\subsubsection{Extensible TLS 1.3 and Hybrid Post-Quantum KEMs}
TLS 1.3 supports the integration of post-quantum KEMs into the handshake through the \texttt{key\_share} extension, a capability not present in TLS 1.2. Post-quantum TLS 1.3 leverages this feature to incorporate post-quantum and hybrid KEMs, such as $\mathsf{X25519MLKEM768}$. Recent IETF working group drafts, including \texttt{draft-ietf-tls-ecdhe-mlkem-03} \cite{ietf-tls-ecdhe-mlkem}, describe these post-quantum primitives and their integration into TLS 1.3, defining algorithm identifiers and usage criteria. The hybrid design combines ECDHE and quantum-safe primitives, which still guarantees security in case one algorithm is broken. This approach enables users to gradually build confidence in post-quantum mechanisms while still retaining the protection offered by the established classical cryptographic schemes. 

\subsubsection{Post-Quantum Authentication and Migration Challenges}
In particular, key exchanges are susceptible to the \textbf{Harvest Now, Decrypt Later} attacks, where a passive attacker can record the encrypted communications in the present with the aim of decrypting them in the future when a cryptographically relevant quantum computer becomes a reality. In contrast, migration of signature schemes is not as urgent because authentication can only happen in real-time, implying that it would require an active quantum attacker to impersonate the other party. However, the scale of migrating to post-quantum signature schemes is significant; thus, standardization efforts to support this transition are already in progress. Internet drafts such as \texttt{draft-ietf-tls-mldsa-latest} \cite{ietf-tls-mldsa} focus on introducing post-quantum authentication mechanisms in TLS 1.3, specifically through composite or pure ML-DSA. Transitioning to post-quantum authentication is more complex, as it requires coordinated upgrades across the Internet’s Public Key Infrastructure—from Root CAs to Intermediate CAs and Leaf Nodes. This systemic requirement slows adoption. Additionally, post-quantum signature schemes introduce a practical drawback in their significantly larger size compared to classical schemes such as ECDSA or RSA. For example, an \texttt{ML-DSA-65} signature is approximately 4 KB, compared to only 64 bytes for ECDSA over \texttt{nistp256}. These increases have contributed to slower adoption due to performance and latency implications. Protocol-level fragmentation, additional round-trip exchanges, and increased channel bandwidth for transmitting post-quantum certificate chains (often $8\mathsf{-}10 \mathsf{KB}$) place pressure on performance-critical networks. 
It's worth mentioning that there have been efforts to redesign WebPKI, considering that the drop-in replacement of classical certificates with post-quantum ones would lead to a noticeable impact on Internet TLS connections. These include the introduction of Merkle-Tree Certificates (MTC) --- \texttt{draft-davidben-tls-merkle-tree-certs} \cite{davidben-tls-merkle-tree-certs}, which benefit from their smaller sizes and enhanced Certificate Transparency. MTC certificates for post-quantum signatures allow for faster verification with fewer signatures to verify.


\section{Identity-Based PKI and TLS Architecture}
In this section, we describe the core components that enable the use of Identity-Based Encryption (IBE) in the de facto application security standard of the web, TLS 1.3. The widespread adoption of TLS in microservices and modern distributed systems motivates the design of IBE-TLS, which aims to provide the same authentication guarantees as standard TLS while eliminating the need for certificates and signature operations. IBE-TLS additionally supports Post-Quantum KEMs (e.g., ID-ML-KEM, X25519MLKEM768).

Realizing identity-based authentication in TLS, however, fundamentally relies on the existence of an underlying IBE-based Public Key Infrastructure (PKI) that defines the space of valid identity strings and associated key extraction mechanisms, and trust anchors. Such an IBE-based PKI is responsible for securely provisioning identity-bound private keys and establishes the trust anchor for all subsequent protocol executions. In our architecture, the Threshold Private Key Generator (T-PKG) acts as the trust anchor of the IBE-based PKI, collectively holding the IBE master secret and securely deriving identity-bound private keys. Conceptually, the T-PKG plays a role analogous to that of a Certification Authority (CA) in an X.509 PKI, serving as the trust anchor of the IBE-based PKI. We therefore first describe the architecture of the IBE-based PKI and T-PKG, followed by its integration into the TLS 1.3 handshake to construct IBE-TLS.

In an identity-based TLS setting, authentication is achieved through key agreement rather than explicit signatures or certificates. Accordingly, the following conditions must be satisfied for successful (mutual) authentication:
\begin{itemize}
  \item Each party is convinced that the peer possesses the private key corresponding to its claimed identity.
  \item Each party is convinced that the peer has observed the same handshake transcript.
  \item Each party is convinced that the peer has derived the same handshake secrets.
\end{itemize}

\subsection{IBE-based PKI Architecture}

We now describe the identity-based public key infrastructure (IBE-based PKI)
used throughout this work. This architecture replaces certificate-based
PKI with an identity-centric trust model, where a Threshold Private Key
Generator (T-PKG) acts as the sole trust anchor for authentication. \footnote{T-PKG is simply an instantiation for the IBE-based PKI, which aims to prevent key escrow, however, deployments and designs may also utilize a centralized PKG, in a private and controlled environment. Our description of IBE-based PKI typically focuses on the T-PKG as the trust anchor, a centralized PKG would follow similar steps, except the key operations --- Setup, Key extraction would not be split amongst multiple parties, and the PKG itself may just be a single entity.}

\subsubsection{Trust Anchor and Public Parameters}

In conventional PKI, trust is rooted in one or more Certificate
Authorities (CAs) whose public keys are distributed to all participants.
In IBE-based PKI, trust is instead rooted in a single set of public system
parameters, referred to as the \emph{master public key} (mpk). The mpk is
globally visible and is used by all parties to derive public keys
corresponding to identities.

The corresponding master secret key is never held by a single entity.
Instead, it is split across multiple T-PKG servers according to a
threshold policy. No individual server can generate identity private
keys on its own; key issuance requires collaboration among a qualified
subset of servers. ~\cite{BendlinKrehbielPeikert2013} develops secure threshold protocols for lattice trapdoor generation, trapdoor delegation, and discrete Gaussian sampling over lattice cosets. These primitives constitute the core building blocks for threshold lattice-based (H)IBE systems, where identity-based private keys are obtained via trapdoor-enabled Gaussian sampling. Their security is proven in the UC framework, with information-theoretic security against adaptive corruptions. The protocols allow for $n$ parties to jointly provide additive shares of a trapdoor matrix $R$ and derive the master public key matrix $A$. Furthermore, they provide a distributed discrete Gaussian sampling protocol which allows for key extraction for a specific identity without reconstructing the master secret. Threshold extraction further mitigates the inherent key escrow problem of IBE by distributing trust across independent servers. These existing threshold protocols therefore suffice to realize a
threshold PKG in practice, allowing identity key extraction to be
distributed across multiple servers without introducing new
cryptographic assumptions.

\subsubsection{Identity-Based Key Lifecycle}

The IBE-based PKI lifecycle consists of the following high-level steps:

\begin{itemize}
  \item \textbf{System initialization:}
  The T-PKG servers jointly establish the system public parameters
  (mpk) and initialize their secret shares. The mpk is distributed to
  all system participants and remains static for the lifetime of the
  trust domain.

  \item \textbf{Identity definition:}
  Identities are defined as structured strings derived from system-level
  identifiers, such as component names, service roles, namespaces, or
  epochs. These identities are deterministic and require no additional
  registration to be usable as public keys.

  \item \textbf{Public key derivation:}
  Any party can derive a public key corresponding to an identity using
  the mpk and the identity string alone. No certificate, directory
  lookup, or interaction with the identity owner is required.

  \item \textbf{Private key issuance:}
  An entity authorized to act under a given identity requests the
  corresponding private key from the T-PKG. Upon successful
  authorization, the T-PKG nodes jointly generate and return the
  identity private key.

  \item \textbf{Key usage and rotation:}
  Identity private keys are used directly in authentication protocols
  (e.g., IBE-TLS). Rotation and revocation are handled by changing the
  identity itself, typically by embedding time- or epoch-based fields.
\end{itemize}

This process eliminates certificate issuance, certificate chains, and
public-key distribution, while preserving explicit trust boundaries
through the T-PKG.

\subsubsection{Identity Registry and Auditability}

Unlike PKI, IBE does not require certificates to bind identities to
public keys. However, operational systems still require visibility into
which identities have been issued and under what conditions, and how long they should be persisted for before getting rotated.

For this purpose, the T-PKG maintains an \emph{identity registry} that
records the issued identities and associated metadata. This registry is
used for auditing, policy enforcement, and incident response, rather
than for key discovery. We describe the identity registry's default fields in Table \ref{tab:identity_registry}.

\begin{table}[h]
\centering
\caption{Identity Registry Maintained by the T-PKG}
\label{tab:identity_registry}
\begin{tabular}{lp{4.2cm}}
\toprule
\textbf{Field} & \textbf{Description} \\
\midrule
Identity$^{*}$ &
Canonical identity string (e.g., service, node, or role) \\

Issuer &
T-PKG trust domain responsible for issuance \\

Authorized principal &
System entity authorized to request this identity \\

Issuance time &
Timestamp of key issuance \\

Validity / epoch &
Time or epoch scope of the identity \\

Status &
Active, expired, or revoked \\
\bottomrule
\end{tabular}
\end{table}

\noindent
$^{*}$The identity itself deterministically defines the corresponding
public key and is therefore not stored for discovery purposes, but for
accountability and auditability.

\subsubsection{Operational Considerations}

The operations of an IBE-based PKI system are close to conventional PKI, which include revocation, trust scoping, internal registries, etc. Despite the similarity, the methodologies are markedly different, because of the avoidance of X.509 certificates, digital signature schemes, and instead relying on identity strings and public/private encryption keypairs.

\paragraph{No public-key distribution.}
Public keys are never generated, stored, or transmitted. Any party can
derive the correct public key from an identity and the mpk at the time
of use.

\paragraph{Just-in-time key generation.}
Identity private keys can be derived only when required and need not be pre-provisioned, unlike X.509 certificates. This naturally aligns with the ephemeral nature of modern workloads and cloud-native deployments. IBE deployments can however, choose to pre-provision keys according to certain epoch periods, allowing for forward secrecy and self-healing capabilities.

\paragraph{Scoped trust domains.}
Multiple T-PKG instances can coexist, each acting as the trust anchor
for a distinct domain, analogous to multiple private CAs in conventional
PKI.

\paragraph{Revocation without CRLs.}
Revocation can achieved by embedding validity constraints into
identities, such as epochs or time windows, rather than maintaining
certificate revocation lists. These epochs are system config, similar to protocols like Messaging Layer Security, allowing the clients to reconstruct the peer's identity without any further protocol interaction.

\paragraph{Reduced blast radius.}
Because the master secret is threshold-distributed, compromise of a
single T-PKG server does not enable arbitrary identity issuance, in
contrast to traditional private CAs.

\subsubsection{Comparison with Certificate-Based PKI}

From a system perspective, the T-PKG plays a role analogous to that of a
private certificate authority, but with different trust and operational
properties. Table~\ref{tab:ibe_pki_vs_ca} summarizes the comparison.

\begin{table*}[h]
\centering
\renewcommand{\arraystretch}{1.3}
\caption{IBE-based PKI vs Certificate-Based PKI}
\label{tab:ibe_pki_vs_ca}
\begin{tabular}{lp{3.8cm}p{3.8cm}}
\toprule
\textbf{Aspect} & \textbf{Certificate-Based PKI} & \textbf{IBE-Based PKI} \\
\midrule
Trust anchor &
CA public key &
Master public key (mpk) \\

Public key form &
X.509 certificate &
Identity string \\

Key distribution &
Certificates transmitted &
Derived on demand \\

Authentication &
Signature verification &
Key decapsulation \\

Revocation &
CRLs / OCSP / short-lived certs &
Identity scoping / epochs \\

Escrow risk &
None &
Mitigated via threshold PKG \\
\bottomrule
\end{tabular}
\end{table*}

\subsection{ID-ML-KEM as the primitive enabling IBE}
This section summarizes the identity-based key encapsulation mechanism
(ID-ML-KEM) proposed in~\cite{id-ml-kem}, which forms the
cryptographic foundation of our protocol. The construction adapts lattice trapdoor sampling techniques in the spirit of the GPV ~\cite{gentry2008trapdoors} framework to build a primitive which takes in parameters and algorithms from the standardized ML-KEM \cite{fips203} and ML-DSA ~\cite{fips204}. The construction uses the Module-NTRU (MNTRU) trapdoor from ModFalcon for key extraction. MNTRU was introduced in the work \cite{CheonKimKimSon2019} which provide an efficient way to generate a trapdoor over MNTRU lattices. 

In particular, we reuse the following algorithms from the paper, for our protocol:

\begin{itemize}
  \item \textbf{Setup} (Algorithm~2): \\
  \emph{Input:} Public system parameters \\
  \emph{Output:} Master public key (\texttt{mpk}), master secret key (\texttt{msk}) \\
  Initializes the IBE system; \texttt{mpk} is distributed as the trust
  anchor, while \texttt{msk} is retained by the issuer (or shared across
  a T-PKG).

  \item \textbf{Extract} (Algorithm~3): \\
  \emph{Input:} Master secret key (\texttt{msk}), identity string
  \texttt{ID} \\
  \emph{Output:} Identity private key \texttt{sk\_ID} \\
  Derives the private key bound to a specific identity; used during
  identity issuance.

  \item \textbf{Encrypt} (Algorithm~4): \\
  \emph{Input:} Master public key (\texttt{mpk}), identity string
  \texttt{ID}, message \texttt{m} \\
  \emph{Output:} Ciphertext \texttt{ct} \\
  Encapsulates key material directly to an identity; used during the
  IBE-TLS handshake for authentication.

  \item \textbf{Decrypt} (Algorithm~5): \\
  \emph{Input:} Identity private key \texttt{sk\_ID}, ciphertext
  \texttt{ct} \\
  \emph{Output:} Message \texttt{m} \\
  Decapsulates the ciphertext; identical to ML-KEM decryption and used
  to recover handshake secrets.
\end{itemize}

\paragraph{Security Properties.}
ID-ML-KEM achieves IND-sID-CPA security under the decisional
Ring-LWE assumption. Security is proven in the selective-identity model,
where the adversary commits to a target identity prior to setup.

\subsection{IBE-based TLS 1.3 Handshake}

With the description of the IBE-based PKI in place, we are now ready to detail IBE-TLS~1.3, which integrates \texttt{ID-ML-KEM} into the TLS~1.3 handshake flow, allowing shared secrets to be derived directly from identity strings. We begin by describing the standard TLS~1.3 handshake, and subsequently the IBE-based modifications.

\subsubsection{Standard TLS 1.3 Certificate-Based Handshake}

In standard TLS~1.3, authentication is achieved using public-key
certificates and digital signatures. The handshake proceeds as follows.

\begin{enumerate}
  \item The client sends a \texttt{ClientHello} message containing the
  supported cipher suites, extensions, and an ephemeral
  \texttt{key\_share} for key exchange.

  \item The server responds with \texttt{ServerHello}, selecting the
  cryptographic parameters and providing its own ephemeral
  \texttt{key\_share}. Both parties derive the handshake traffic keys.

  \item The server then sends \texttt{EncryptedExtensions}, followed by
  its \texttt{Certificate} message containing the certificate chain,
  and a \texttt{CertificateVerify} message with a digital signature over
  the handshake transcript.

  \item The server concludes its authentication by sending a
  \texttt{Finished} message, which provides key confirmation.

  \item The client verifies the certificate chain and signature, and in
  the mutual TLS case responds with its own \texttt{Certificate},
  \texttt{CertificateVerify}, and \texttt{Finished} messages.
\end{enumerate}

\subsubsection{IBE-TLS Handshake Overview}

The TLS handshake is crucial for establishing the shared secrets and keys for further application data exchange. It comprises a set of messages, an internal key schedule, and further protocol-level considerations. Our modification of TLS 1.3 (PQ) to introduce IBE-based schemes (ID-ML-KEM) doesn't aim to add new messages or extra round-trips, but to shorten and drop messages that are deemed unnecessary after the addition of IBE. Since IBE removes the need for signature schemes and X.509 certificates, we remove the messages that carry them — \texttt{Certificate, CertificateRequest}, and
\texttt{CertificateVerify}. Other messages, such as \texttt{ClientHello, ServerHello, ClientFinished, ServerFinished}, from TLS 1.3 are reused, but with important additions for making IBE-based authentication possible. An overview of the used handshake flow and its considerations with IBE is described in the Figure ~\ref{fig:ibe-tls-handshake}.

\begin{figure*}[t]
\centering
\begin{tikzpicture}[
    node distance=1.6cm,
    box/.style={
        rectangle, draw, thick, rounded corners,
        minimum width=3.2cm, minimum height=1.0cm,
        align=center
    },
    msg/.style={->, thick},
]
\node[box] (client) {Client\\$\mathsf{ID}_C$};
\node[box, right=7.5cm of client] (server) {Server\\$\mathsf{ID}_S$};

\draw[msg]
([yshift=-0.8cm]client.south) --
([yshift=-0.8cm]server.south)
node[midway, above, align=center] {
\texttt{ClientHello}\\
\footnotesize
X25519 key share $eph_C$\\
ID-ML-KEM-768 ciphertext $ct_{\mathsf{ID}_S}$
};
\draw[msg]
([yshift=-2.8cm]server.south) --
([yshift=-2.8cm]client.south)
node[midway, above, align=center] {
\texttt{ServerHello}\\
\footnotesize
X25519 key share $eph_S$\\
ID-ML-KEM-768 ciphertext $ct_{\mathsf{ID}_C}$
};

\draw[msg]
([yshift=-3.5cm]server.south) --
([yshift=-3.5cm]client.south)
node[midway, above] {\texttt{EncryptedExtensions}};

\node[below=4.5cm of client, align=left, text width=5.4cm] {
{\footnotesize
\textbf{Client side:}\\[4pt]
$\bullet$ Derive ephemeral secret from X25519\\
$\bullet$ Decapsulate $ct_{\mathsf{ID}_C}$ using $sk_{\mathsf{ID}_C}$\\
$\bullet$ Derive handshake secret
}};

\node[below=4.5cm of server, align=left, text width=5.4cm] {
{\footnotesize
\textbf{Server side:}\\[4pt]
$\bullet$ Derive ephemeral secret from X25519\\
$\bullet$ Decapsulate $ct_{\mathsf{ID}_S}$ using $sk_{\mathsf{ID}_S}$\\
$\bullet$ Derive handshake secret
}};

\draw[msg]
([yshift=-7.4cm]server.south) --
([yshift=-7.4cm]client.south)
node[midway, above] {\texttt{ServerFinished}};

\draw[msg]
([yshift=-8cm]client.south) --
([yshift=-8cm]server.south)
node[midway, above] {\texttt{ClientFinished}};

\end{tikzpicture}
\caption{Overview of a TLS~1.3 handshake with identity-based authentication
using ID-ML-KEM-768 and X25519. Identity-based encapsulation replaces
certificate-based authentication, while the TLS~1.3 key schedule and
Finished messages provide key confirmation and implicit authentication.}
\label{fig:ibe-tls-handshake}
\end{figure*}

A brief idea (using ID-ML-KEM) is described below, which we will expand in later sections:

\begin{itemize}

\item The client, using a dedicated service discovery mechanism, discovers the server’s identity and encapsulates to it. The resulting ciphertext is then included in their ClientHello message, in addition to the existing field(s); the shared secret is undisclosed. Note that the existence of a service discovery mechanism is not always necessary, and in scenarios where the identity is just the domain name, it can be used instead, which is similar to the use of \texttt{ServerNameIndication} extension in TLS 1.3.

Addition: In private environments, the client may also send their own identity, since those may necessitate mutual authentication. If the client doesn’t share it, the server can request it in a subsequent message. In public domains, though, this is discouraged, since it would leak the client’s identity, unless Encrypted ClientHello (ECH) is used.

\item The server, if the correct intended peer, will be able to decapsulate the sent ciphertext with overwhelming probability, if they hold the secret key to that identity. The resulting shared secret will be kept at their disposal, with the server then encapsulating the client’s ephemeral key share and their identity string (if sent and mutual authentication is mandatory). The resulting shared secrets — ephemeral key share, client’s identity, and server’s identity- will form the base for the derivation of all other secrets, such as Application secrets, Traffic secrets, etc.

\item The server could send a message similar to HelloRetryRequest from TLS 1.3, where it asks for the client’s identity if not received beforehand, prompting the
client to resend the ClientHello with the required information. The handshake then continues as described above.

\item The rest of the TLS flow continues, with the Server responding with ServerFinished — a MAC over the entire transcript, and the Client doing the same. Correct MAC tags guarantee that both peers derived the same keys and saw the same transcript. It also proves possesion of the secret key corresponding to that identity, thus, satisfying all the 3 conditions laid out previously for mutual authentication.

\end{itemize}

\subsubsection{Message Integration}

In this section, we describe how the IBE components—namely the encapsulated peer identity, the client’s own identity, and the selected IBE scheme—are integrated into the TLS~1.3 handshake. To achieve this, we introduce two new TLS extensions: \texttt{ibe\_identity\_auth}, which carries the ciphertext obtained after encapsulating the peer’s identity under an IBE scheme, and \texttt{ibe\_identity}, which is used by a party to convey its own identity to the peer.

TLS extensions are typed, length-delimited, and negotiable parameters that are attached to handshake messages in order to extend the core protocol without modifying the base protocol. The TLS~1.3 specification defines several such extensions, including \texttt{key\_share}, \texttt{signature\_algorithms}, and \texttt{server\_name}. Each extension is identified by an \texttt{ExtensionType}, an unsigned integer that determines how the associated data is interpreted. The extension contents are carried inside a generic \texttt{Extension} structure, defined as follows:

\begin{verbatim}
struct {
    ExtensionType extension_type;
    opaque extension_data<0..2^16-1>;
} Extension;
\end{verbatim}
For example, the \texttt{key\_share} extension is represented as:
\begin{verbatim}
Extension{
    extension_type = 51,
    extension_data = encode (
        KeyShareClientHello{
            client_shares = [
                KeyShareEntry{
                    NamedGroup: X25519MLKEM768,
                    key_exchange = 
                    <client_pub_key>
                }
            ]
        }
    )
}
\end{verbatim}
where, \texttt{encode(.)} is an implementation specific function, responsible for encoding the extension data to the \textbf{opaque} object defined in the TLS presentation language.

We reuse this structure to define the two IBE-specific extensions described below.

The \texttt{ibe\_identity\_auth} extension is used during the negotiation phase of the handshake to convey the encapsulated identity of the peer, along with an identifier for the IBE scheme in use. This extension is carried in the \texttt{ClientHello} and \texttt{ServerHello} messages, and thus participates directly in handshake negotiation.

The \texttt{extension\_data} field of \texttt{ibe\_identity\_auth} is structured as follows:

\begin{verbatim}
struct {
    uint16 ibe_scheme_id;
    opaque encapsulated_identity<1..2^16-1>;
} IBEIdentityAuth;
\end{verbatim}

An example instantiation of this extension inside a TLS handshake message is shown below:

\begin{verbatim}
Extension {
    extension_type = ibe_identity_auth,
    extension_data = IBEIdentityAuth {
        ibe_scheme_id = 0x0001,
        encapsulated_identity = 
        <IBE-enc(identity_peer)>
    }
}
\end{verbatim}
where the \texttt{id\_scheme\_id = 0x0001} indicates ID-ML-KEM.

The \texttt{ibe\_identity} extension is used to convey a party’s own identity in clear form and serves as authentication-related metadata. This extension is sent in the ClientHello, for scenarios which require mutual authentication --- particularly, private environments. 
In scenarios where the server requires additional identity material, wishes to trigger re-encapsulation, or receive client's identity (if not sent previously), it MAY request the client’s \texttt{ibe\_identity\_auth} extension via a \texttt{HelloRetryRequest}, prior to
\texttt{ServerHello}, requesting the client’s \texttt{ibe\_identity\_auth} extension.

The structure of the \texttt{ibe\_identity} extension is defined as follows:

\begin{verbatim}
struct {
    opaque identity<1..2^16-1>;
} IBEIdentity;
\end{verbatim}

An example instantiation of the \texttt{ibe\_identity} extension is shown below:

\begin{verbatim}
Extension {
    extension_type = ibe_identity,
    extension_data = IBEIdentity {
        identity = "client@example.com"
    }
}
\end{verbatim}

This design follows the TLS~1.3 principle of separating negotiation parameters from authentication material. The \texttt{ibe\_identity\_auth} extension contributes to the handshake transcript and is authenticated by the Finished messages, ensuring agreement on the encapsulated identity and IBE scheme. In contrast, the \texttt{ibe\_identity} extension is bound to the authentication phase of the handshake and is verified implicitly
through successful identity-based decapsulation and transcript
authentication. This allows linking the subsequent session with the transmitted identity.

\subsubsection{Key Schedule Integration}

Firstly, we provide an overview of the standard TLS 1.3 key schedule, thereafter presenting a construction which integrates the IBE KEM into it. We later analyze its security properties.

The standard TLS 1.3 key schedule uses a chain of HKDF — Extract and Expand calls, which are used to derive the specific keys, based on the present context and handshake state. The entropy arises from either a pre-shared key mixed with a secret derived from the ephemeral key share — $\mathsf{eph_s}$ or only $\mathsf{eph_s}$.

The Figure ~\ref{fig:tls13-key-schedule-schematic} provides the TLS 1.3 key schedule overview, divided into client and server side boxes for clarity.

\begin{figure*}[h]
\centering
\setlength{\tabcolsep}{8pt}
\renewcommand{\arraystretch}{1.15}
\small
\begin{tabular}{|p{0.45\textwidth}|p{0.45\textwidth}|}
\hline
\multicolumn{2}{|c|}{\textbf{TLS 1.3 Key Schedule (Schematic Overview)}} \\
\hline
\textbf{Client} & \textbf{Server} \\
\hline

$\mathsf{early\_secret}
 = \mathsf{HKDF\text{-}Extract}(0, \mathsf{PSK})$
&
$\mathsf{early\_secret}
 = \mathsf{HKDF\text{-}Extract}(0, \mathsf{PSK})$
\\[0.2em]

$\begin{aligned}[t]
\mathsf{derived\_secret} &=
\mathsf{HKDF\text{-}Expand\text{-}Label}(
\mathsf{early\_secret},\\
&\qquad \texttt{"derived"}, \epsilon)
\end{aligned}$
&
$\begin{aligned}[t]
\mathsf{derived\_secret} &=
\mathsf{HKDF\text{-}Expand\text{-}Label}(
\mathsf{early\_secret},\\
&\qquad \texttt{"derived"}, \epsilon)
\end{aligned}$
\\[0.2em]

$\mathsf{handshake\_secret}
 = \mathsf{HKDF\text{-}Extract}(
 \mathsf{derived\_secret}, \mathsf{eph_s})$
&
$\mathsf{handshake\_secret}
 = \mathsf{HKDF\text{-}Extract}(
 \mathsf{derived\_secret}, \mathsf{eph_s})$
\\[0.2em]

$\begin{aligned}[t]
\mathsf{client\_hs\_traffic\_secret} &=
\mathsf{HKDF\text{-}Expand\text{-}Label}\\(
\mathsf{handshake\_secret},
&\ \texttt{"c hs traffic"}, \mathsf{th}_1)
\end{aligned}$
&
$\begin{aligned}[t]
\mathsf{server\_hs\_traffic\_secret} &=
\mathsf{HKDF\text{-}Expand\text{-}Label}\\(
\mathsf{handshake\_secret},
&\ \texttt{"s hs traffic"}, \mathsf{th}_1)
\end{aligned}$
\\[0.2em]

$\begin{aligned}[t]
\mathsf{client\_finished\_key} &=
\mathsf{HKDF\text{-}Expand\text{-}Label}\\(
\mathsf{client\_hs\_traffic\_secret},
&\ \texttt{"finished"}, \epsilon)
\end{aligned}$
&
$\begin{aligned}[t]
\mathsf{server\_finished\_key} &=
\mathsf{HKDF\text{-}Expand\text{-}Label}\\(
\mathsf{server\_hs\_traffic\_secret},
&\ \texttt{"finished"}, \epsilon)
\end{aligned}$
\\[0.4em]

$\begin{aligned}[t]
\mathsf{derived\_secret} &=
\mathsf{HKDF\text{-}Expand\text{-}Label}(
\mathsf{handshake\_secret},\\
&\qquad \texttt{"derived"}, \epsilon)
\end{aligned}$
&
$\begin{aligned}[t]
\mathsf{derived\_secret} &=
\mathsf{HKDF\text{-}Expand\text{-}Label}\\(
\mathsf{handshake\_secret},\\
&\qquad \texttt{"derived"}, \epsilon)
\end{aligned}$
\\[0.2em]

$\mathsf{master\_secret}
 = \mathsf{HKDF\text{-}Extract}(\mathsf{derived\_secret}, 0)$
&
$\mathsf{master\_secret}
 = \mathsf{HKDF\text{-}Extract}(\mathsf{derived\_secret}, 0)$
\\[0.2em]

$\begin{aligned}[t]
\mathsf{client\_app\_traffic\_secret}_0 &=
\mathsf{HKDF\text{-}Expand\text{-}Label}\\(
\mathsf{master\_secret},
&\ \texttt{"c ap traffic"}, \mathsf{th}_2)
\end{aligned}$
&
$\begin{aligned}[t]
\mathsf{server\_app\_traffic\_secret}_0 &=
\mathsf{HKDF\text{-}Expand\text{-}Label}\\(
\mathsf{master\_secret},
&\ \texttt{"s ap traffic"}, \mathsf{th}_2)
\end{aligned}$
\\

\hline
\end{tabular}

\caption{Schematic representation of the TLS~1.3 key schedule. Context strings
ensure domain separation, while transcript hashes $\mathsf{th}_1$ and
$\mathsf{th}_2$ bind derived secrets to the handshake state.}
\label{fig:tls13-key-schedule-schematic}
\end{figure*}

IBE-TLS, on the other hand, modifies the initial key schedule steps, particularly until the formation of the handshake secret. The shared secrets are formed as follows:
\begin{itemize}

\item \textbf{ClientHello:}
The client generates an ephemeral key pair
$(\mathsf{sk}_{c}^{\mathsf{eph}}, \mathsf{pk}_{c}^{\mathsf{eph}})$.
Using the server identity $\mathsf{ID}_s$ and the system master public
key $\mathsf{mpk}$, the client performs an identity-based encapsulation:
\[
  (\mathsf{ct}_s, \mathsf{ss}_s) \leftarrow
  \mathsf{ID\text{-}ML\text{-}KEM.Encaps}(\mathsf{mpk}, \mathsf{ID}_s).
\]
The client sends its ephemeral public key $\mathsf{pk}_{c}^{\mathsf{eph}}$
and the ciphertext $\mathsf{ct}_s$ in the \texttt{ClientHello} message.

\item \textbf{ServerHello:}
Upon receiving the \texttt{ClientHello}, the server generates its own
ephemeral key pair
$(\mathsf{sk}_{s}^{\mathsf{eph}}, \mathsf{pk}_{s}^{\mathsf{eph}})$ and
derives the ephemeral shared secret
\[
  \mathsf{eph} \leftarrow
  \mathsf{KEX}(\mathsf{sk}_{s}^{\mathsf{eph}}, \mathsf{pk}_{c}^{\mathsf{eph}}).
\]
The server decapsulates the identity-based ciphertext using its identity
secret key $\mathsf{sk}_{\mathsf{ID}_s}$:
\[
  \mathsf{ss}_s \leftarrow
  \mathsf{ID\text{-}ML\text{-}KEM.Decaps}(\mathsf{sk}_{\mathsf{ID}_s}, \mathsf{ct}_s).
\]

For mutual authentication, the server additionally encapsulates with
respect to the client identity $\mathsf{ID}_c$:
\[
  (\mathsf{ct}_c, \mathsf{ss}_c) \leftarrow
  \mathsf{ID\text{-}ML\text{-}KEM.Encaps}(\mathsf{mpk}, \mathsf{ID}_c),
\]
and sends $\mathsf{ct}_c$ in the \texttt{ServerHello}.
The client recovers $\mathsf{ss}_c$ via
\[
  \mathsf{ss}_c \leftarrow
  \mathsf{ID\text{-}ML\text{-}KEM.Decaps}(\mathsf{sk}_{\mathsf{ID}_c}, \mathsf{ct}_c).
\]
The client derives the ephemeral shared secret upon receiving the
\texttt{ServerHello}:
\[
  \mathsf{eph} \leftarrow
  \mathsf{KEX}(\mathsf{sk}_{c}^{\mathsf{eph}}, \mathsf{pk}_{s}^{\mathsf{eph}}).
\]

\end{itemize}

At the conclusion of the hello exchange, both parties hold the same set
of shared secrets
\[
  \{\mathsf{eph}, \mathsf{ss}_s, \mathsf{ss}_c\},
\]
where $\mathsf{ss}_c$ is present only in the mutual-authentication
setting.

\paragraph{Handshake Secret Derivation.}
After the completion of the \texttt{ClientHello} and \texttt{ServerHello}
exchange, both parties possess a common set of shared secrets derived
from the ephemeral key exchange and the identity-based encapsulations.
These secrets are combined to derive the TLS~1.3 handshake secret.

Let $\mathsf{derived}$ denote the output of the standard TLS~1.3
\texttt{"derived"} expansion from the early secret.
The handshake secret is computed as:
\[\begin{aligned} 
  \mathsf{handshake\_secret} \leftarrow
  \mathsf{HKDF\text{-}Extract}\\(
    \mathsf{derived},
    \mathsf{eph} \,\|\, \mathsf{ss}_s \,\|\, \mathsf{ss}_c
),
\end{aligned}\]
where $\|$ denotes concatenation, and $\mathsf{ss}_c$ is omitted in the
unilateral-authentication setting.

The resulting $\mathsf{handshake\_secret}$ replaces the Diffie--Hellman
shared secret used in standard TLS~1.3. All subsequent key derivations,
including the client and server handshake traffic secrets, finished
keys, master secret, and application traffic secrets, follow the
unmodified TLS~1.3 key schedule and context string definitions.

In the following section, we discuss the security properties and threat model of the proposed protocol and analyze the security guarantees it offers.

\subsection{Security Analysis}

\subsubsection{Implicit Authentication}

Authentication is achieved as follows. The server can only derive $ss_S$ if it possesses $sk_{ID_S}$, which can only be obtained from the PKG for identity $ID_S$. Successful completion of the handshake (verified via the \texttt{Finished} MAC) proves that the server performed correct decapsulation, implying possession of the correct secret key. This provides equivalent authentication to certificate-based TLS, where the server proves possession of the private key corresponding to its certified public key (and the claimed identity). The same stands true for client authentication.

\subsubsection{Mutual Authentication}

Mutual authentication follows symmetrically. The client must possess $sk_{ID_C}$ to decapsulate $ct_C$ and derive the correct shared key. Both parties authenticate each other simultaneously through the combined key derivation.

\subsubsection{Forward Secrecy}

Forward secrecy is achieved by incorporating ephemeral key exchange into
the handshake procedure. The use of identity-based encryption primitives
does not alter the standard TLS key exchange, which is ephemeral by
default and includes ECDHE or hybrid ML-KEM key shares. 

In addition to the standard key shares, identity-based encapsulations
are included to provide authentication. The final session key is derived
by combining the ephemeral shared secret with the identity-based shared
secrets:
\[
K = \mathsf{KDF}(ss_{\text{ephemeral}} \| ss_S \| ss_C )
\]
As a result, compromise of long-term identity secret keys alone does not
enable recovery of past session keys, provided the ephemeral secrets
remain uncompromised.

\subsection{ML-KEM-based IBE-TLS Compared to Post-Quantum TLS}

We compare the proposed ML-KEM-based IBE-TLS handshake with standard
certificate-based post-quantum TLS~1.3. The comparison focuses on
mechanical differences in the handshake, authentication-related
cryptographic operations, and authentication-specific bandwidth
overhead. The TLS~1.3 key schedule, record layer, and symmetric
cryptography remain unchanged; all differences arise solely from the
authentication mechanism.

Table~\ref{tab:handshake-messages} lists the handshake messages exchanged
during mutual authentication. In IBE-TLS, certificate-related messages
are omitted entirely, as authentication is achieved implicitly through
successful identity-based key decapsulation rather than explicit
certificate and signature verification.

\begin{table*}[t]
\centering
\caption{TLS~1.3 Handshake Message Comparison}
\label{tab:handshake-messages}
\rowcolors{2}{lightgrayrow}{white}
\begin{tabular}{>{\raggedright\arraybackslash}p{5.5cm}cc}
\toprule
\textbf{Handshake Message} & \textbf{Cert-based PQ-TLS} & \textbf{ID-based IBE-TLS} \\
\midrule
ClientHello               & \cmark & \cmark \\
ServerHello               & \cmark & \cmark \\
EncryptedExtensions       & \cmark & \cmark \\
Certificate               & \cmark & \xmark \\
CertificateRequest (mTLS) & \cmark & \xmark \\
CertificateVerify         & \cmark & \xmark \\
Finished (Server)         & \cmark & \cmark \\
Finished (Client)         & \cmark & \cmark \\
\bottomrule
\end{tabular}
\end{table*}

\subsubsection{Authentication-Related Cryptographic Operations}

Table~\ref{tab:crypto-ops} summarizes the asymmetric cryptographic
operations performed during a mutually authenticated handshake. Counts
are per handshake and assume a post-quantum signature scheme such as
ML-DSA for certificate-based TLS 
\footnote{This includes the standard TLS ephemeral key exchange in KEM encaps and decaps}
. 

\begin{table*}[t]
\centering
\caption{Authentication-Related Cryptographic Operations (Mutual TLS) 
}
\label{tab:crypto-ops}
\rowcolors{2}{lightgrayrow}{white}
\begin{tabular}{>{\raggedright\arraybackslash}p{7cm}cc}
\toprule
\textbf{Operation} & \textbf{Cert-based PQ-TLS} & \textbf{ID-based IBE-TLS} \\ 
\midrule
KEM encapsulation                        & 1 & 3 \\
KEM decapsulation                        & 1 & 3 \\
Post-quantum signature generation        & 2 & 0 \\
Post-quantum signature verification      & 4 & 0 \\
Certificate parsing and validation        & Required & Not required \\
Certificate chain path validation         & Required & Not required \\
Identity public key derivation            & Not required & Required \\
\bottomrule
\end{tabular}
\end{table*}

Certificate-based TLS incurs additional cost from post-quantum signature
generation and verification as well as certificate parsing and chain
validation, all of which lie on the handshake critical path. IBE-TLS
eliminates these steps entirely and replaces them with deterministic
public key derivation from identity strings, consisting primarily of
hashing and linear operations.

\subsubsection{Authentication Bandwidth Overhead}

Table~\ref{tab:bandwidth} compares authentication-related handshake
bandwidth for certificate-based post-quantum TLS and IBE-TLS.

\begin{table*}[t]
\centering
\caption{Authentication-Related Handshake Bandwidth Overhead (Single-side authentication)}
\label{tab:bandwidth}
\rowcolors{2}{lightgrayrow}{white}
\begin{tabular}{>{\raggedright\arraybackslash}p{7cm}cc}
\toprule
\textbf{Component} & \textbf{Cert-based PQ-TLS} & \textbf{ID-based IBE-TLS} \\
\midrule
Certificate chain(s)                & 8--15 KB & 0 KB \\
CertificateVerify signatures        & 3--6 KB  & 0 KB \\
Identity-based authentication data  & 0 KB     & $\sim$5 KB \\
\midrule
\textbf{Total authentication data}  & \textbf{11--21 KB} & \textbf{$\sim$5 KB} \\
\bottomrule
\end{tabular}
\end{table*}

Identity-based authentication instantiated using ID-ML-KEM introduces
ciphertexts on the order of several kilobytes. As a result, the
bandwidth reduction achieved by IBE-TLS is more modest than in classical
IBE constructions. However, certificate chains and transcript
signatures—which together dominate authentication bandwidth in
post-quantum TLS—are eliminated entirely. Unlike certificates, the
ID-ML-KEM ciphertext size is fixed and independent of certificate
hierarchy depth or validation policy.

\section{Performance Analysis}

This section analyzes the performance implications of the proposed
identity-based TLS handshake relative to certificate-based
post-quantum TLS. Rather than relying on aggregate handshake size alone,
we relate performance directly to the authentication mechanisms removed
or replaced by IBE-TLS.

\subsection{Bandwidth Implications}

In certificate-based post-quantum TLS, authentication bandwidth is
dominated by certificate chains and \texttt{CertificateVerify}
signatures, which are transmitted on every handshake and scale with
certificate structure and signature parameters. In contrast, IBE-TLS
replaces these artifacts with a fixed number of identity-based KEM
ciphertexts embedded in the \texttt{key\_share} exchange. As a result,
authentication bandwidth becomes bounded and predictable across
handshakes.

\subsection{Computational Implications}

Certificate-based TLS performs multiple post-quantum signature
verifications and certificate validation steps during authentication.
These operations (such as SAN verification, chain validation) are computationally expensive and sensitive to
certificate size and hierarchy depth. IBE-TLS removes these steps and
reduces authentication to a fixed sequence of KEM operations and
identity-derived public key computation, yielding a more uniform
computational profile.

\subsection{Latency Implications}

The latency implications of IBE-TLS follow from the removal of
certificate-dependent authentication steps rather than changes to the
TLS~1.3 handshake structure. In certificate-based deployments,
handshake completion depends on signature verification and certificate
validation on the critical path. By replacing these with a fixed number
of KEM operations, IBE-TLS yields more predictable authentication
latency, particularly in environments with frequent connection
establishment.

\section{Integration of IBE to Application Workloads}

We begin by describing how an Identity-Based Encryption (IBE)–enabled TLS system, alongside conventional PKI, can be integrated into cloud-native application workloads deployed on Kubernetes. Our approach focuses on securing service-to-service communication at the application layer and operates independently of the Kubernetes control plane. We present a concrete architecture and demonstrate the feasibility of the proposed system using \textbf{QORE}, a post-quantum 5G Core that currently relies on PKI with post-quantum certificates.
The high-level objectives of this work include:
\begin{itemize}
    \item Replace conventional PKI based on Certificate Authorities with an IBE-based PKI that uses a  Private Key Generator (T-PKG/PKG).
    \item Use the proposed PKI system to provide end-to-end, mutually authenticated IBE-TLS communication between constituent services.
    \item Provide post-quantum security for microservice communication using \texttt{ID-ML-KEM}.
    \item Reduce certificate-induced operational overhead through the use of an IBE-based PKI.
    \item Reduce bandwidth and computational costs, and potentially improve latency and throughput, by using IBE-TLS in place of (PQ)TLS~1.3.
    \item Act as a drop-in replacement that preserves existing communication semantics, security guarantees, and deployment structure.
\end{itemize}

The goals stated above align with the demands of modern, latency-sensitive applications, for which the 5G Core serves as a representative example.

\subsection{Identity Namespace Design}

Kubernetes provides built-in identity concepts including namespaces, service accounts, and pod labels. We define microservice identities using the structure:
\[
ID = \text{cluster-name} \| \text{namespace} \| \text{service} \| \text{epoch}
\]

The cluster name distinguishes different Kubernetes clusters in multi-cluster deployments. The namespace provides logical isolation, typically corresponding to teams or applications. The service field identifies the specific microservice. The epoch is a version identifier enabling key rotation, formatted as a timestamp or monotonic counter.
Alternatively, one can prefer a format where identitties are described in a manner similar to Kubernetes logical service identifiers (discussed previously). 

For example:
\begin{verbatim}
prod-us-west.payments.checkout-api.20250101
\end{verbatim}

This identity structure leverages existing Kubernetes concepts without requiring new identity management infrastructure.

\subsection{T-PKG Deployment}
In Kubernetes application workloads, PKI is often used to implement
mutual authentication between the communicating services and guarantee a
zero-trust architecture. Deployments routinely use Certificate
Authorities setup by service meshes (such as Istio) or use an external
enterprise-level CA (such as EJBCA). These manage the entire certificate
lifecycle for the services, including issuance, registration,
revocation, and provides OCSP and CRL responders. These CAs are often
times deployed as a separate Kubernetes namespace secured with
hardware-backed secret keys. They usually provide a common endpoint for
the i) Registration Authority ii) Validation Authority iii) Certificate
Authority, etc. These operate as pods, having persistent volume stores
and databases, which are essential for their operations.

In addition to application workloads, Kubernetes itself relies on a
private PKI for securing control-plane and node communication. Core
components such as the API server, kubelets, and etcd authenticate using
certificates issued by cluster-local CAs, following the same operational
model of centralized issuance, persistent state, and certificate-based
trust distribution.

A typical deployment looks like:
\small
\begin{verbatim}
$ kubectl get pods -n pki-system

ca-server-0            1/1   Running   0   120d
ra-server              1/1   Running   2   120d
ocsp-responder         1/1   Running   1   120d
pki-database-0         1/1   Running   0   120d
\end{verbatim}
\normalsize
We reuse the same framework, but replace the Certificate Authority with a
T-PKG, where the participating nodes of the T-PKG operate as separate
services, while the external API procedures for registration, issuance
and validation remain the same, except for the replacement of
certificates with identity-backed private keys.

In this setting, the T-PKG exposes a single registration and issuance
endpoint, analogous to a conventional CA service. Internally, this
endpoint coordinates a set of T-PKG nodes/pods, each holding a share of the
master secret. Upon an authorized request, the endpoint collects partial
responses from a threshold of T-PKG nodes and combines them ephemerally
in memory to construct the identity private key. The assembled key is
returned to the client and immediately erased, while no private key
material is persisted in storage.

A T-PKG-based deployment follows the same operational structure, with a
single externally visible registration endpoint coordinating multiple
internal services.
\small
\begin{verbatim}
$ kubectl get pods -n ibe-pki

tpkg-register            1/1   Running   2   90d
tpkg-node-1              1/1   Running   1   90d
tpkg-node-2              1/1   Running   3   90d
tpkg-node-3              1/1   Running   0   90d
identity-registry-db     1/1   Running   0   90d

$ kubectl get svc -n ibe-pki

tpkg-register    ClusterIP   10.96.120.12   443/TCP
tpkg-internal    ClusterIP   None           8443/TCP
\end{verbatim}
\normalsize

\subsection{Key Lifecycle Management}

In a way that is analogous to certificate lifecycle management systems in
Kubernetes—which automate certificate issuance, rotation, revocation (e.g., cert-manager \cite{certmanager}),
and provisioning—we describe a key lifecycle management mechanism for
identity-based cryptographic keys. This mechanism is responsible for
securely provisioning, rotating, and retiring identity private keys
associated with Kubernetes services and pods, while integrating
directly with existing Kubernetes identity and authorization
primitives.

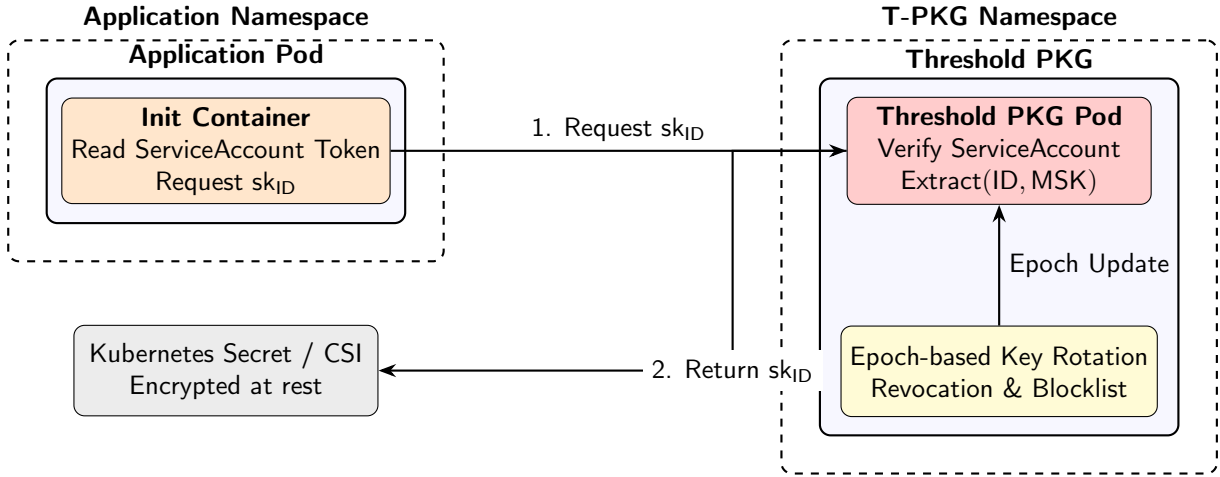
\begin{figure*}
    \centering
\begin{tikzpicture}[
  font=\sffamily,
  node distance=1.6cm,
  ,
  box/.style={
    draw,
    rounded corners,
    align=center,
    minimum width=4cm,
    minimum height=1.2cm,
    fill=gray!5
  },
  pod/.style={
    draw,
    thick,
    rounded corners,
    inner sep=10pt,
    fill=blue!3
  },
  ns/.style={
    draw,
    dashed,
    thick,
    rounded corners,
    inner sep=14pt
  },
  arrow/.style={
    ->,
    thick,
    >=Stealth
  },
 double_arrow/.style = {<->,
 thick,
 >=Stealth}
]


\node[box, fill=orange!20] (init) {};
\node[box, fill=gray!15] (secret) {};

\node[pod, fit=(init)(secret),
label=above:{\textbf{Application Pod}}] (app-pod) {};

\node[ns, fit=(app-pod),
label=above:{\textbf{Application Namespace}}] (app-ns) {};

\node[box, fill=orange!20] (init) {
\textbf{Init Container}\\
Read ServiceAccount Token\\
Request $\mathsf{sk}_{\mathsf{ID}}$
};

\node[box, fill=gray!15, below=of init] (secret) {
Kubernetes Secret / CSI\\
Encrypted at rest
};


\node[box, fill=red!20, right=6cm of init] (pkg) {};
\node[box, fill=yellow!20, below=of pkg] (rotation) {};

\node[pod, fit=(pkg)(rotation),
label=above:{\textbf{Threshold PKG}}] (pkg-pod) {};

\node[ns, fit=(pkg-pod),
label=above:{\textbf{T-PKG Namespace}}] (pkg-ns) {};

\node[box, fill=red!20, right=6cm of init] (pkg) {
\textbf{Threshold PKG Pod}\\
Verify ServiceAccount\\
$\mathsf{Extract}(\mathsf{ID}, \mathsf{MSK})$
};

\node[box, fill=yellow!20, below=of pkg] (rotation) {
Epoch-based Key Rotation\\
Revocation \& Blocklist
};

\draw[arrow] (init.east) -- node[above]{1. Request $\mathsf{sk}_{\mathsf{ID}}$} (pkg.west);

\draw[arrow] (pkg.west) -- ++(-1.5,0)
  |- node[midway, fill=white]{2. Return $\mathsf{sk}_{\mathsf{ID}}$} (secret.east);

\draw[arrow] (rotation.north) -- node[right]{Epoch Update} (pkg.south);


\end{tikzpicture}
\caption{Key lifecycle management with T-PKG}
\label{fig:keylifecycle}
\end{figure*}

Figure \ref{fig:keylifecycle} shows the Key Lifecycle management for a Kubernetes Pod.

\subsubsection{Initial Key Provisioning}

When a new pod is created, an Init Container~\cite{kubernetes-init-containers-docs} runs before the application container starts. This Init Container performs the following steps:

\begin{algorithm}
\caption{Pod Identity Key Bootstrap}
\begin{algorithmic}[1]
\State Read service account token from mounted volume
\State Construct identity $ID$ from pod metadata
\State Request $sk_{ID}$ from T-PKG service endpoint using service account for authentication 
\State Store $sk_{ID}$ in shared volume or environment variable
\State Exit, allowing main application container to start
\end{algorithmic}
\end{algorithm}

A pod is issued an identity key after successful Kubernetes-native authentication. For this, the algorithm uses the pod’s ServiceAccount token, which is a signed JWT issued by the Kubernetes API server~\cite{kubernetes-service-account-docs}. Upon receiving a key issuance request, the T-PKG verifies the token either by validating its signature against the API server’s public keys or by invoking the TokenReview API. This builds an inherent trust chain, where the T-PKG trusts the Kubernetes API server, and only after successful verification, the identity key is granted, ensuring the pod is both authenticated and authorized under the cluster’s RBAC policies. The key is shared over the secure IBE-TLS connection, where the pod encapsulates the T-PKG's identity (e.g., tpkg-node1) for server-side authentication. 
The ServiceAccount identities are then mapped to the corresponding cryptographic identities, which are internally logged. This builds an inherent trust chain, where the T-PKG trusts the Kubernetes API server. 

\subsubsection{Key Rotation}

Key rotation is triggered by incrementing the epoch field in the identity. This can occur on a fixed schedule (e.g., daily) or in response to security events. When the epoch changes, new pods automatically receive new keys during bootstrap. Existing pods are gradually rotated through standard Kubernetes deployment updates.

During a rotation period, both old and new epoch keys could be valid, allowing graceful transition. The transition window is configurable based on deployment velocity and security requirements.



\subsubsection{Revocation}

Revocation in identity-based systems differs from certificate-based PKI. Instead of maintaining revocation lists, we use short-lived epochs. An identity with epoch $e$ is automatically invalidated when the system transitions to epoch $e+1$. Emergency revocation can be achieved by triggering an immediate epoch increment for affected services.

Additionally, the T-PKG can maintain a blocklist of explicitly revoked identities, refusing to issue keys for listed identities regardless of epoch. This provides defense-in-depth for compromised pods.

\subsection{Integration with Kubernetes Secrets}

Identity keys are stored as Kubernetes Secrets. The Secret object contains:

\begin{verbatim}
apiVersion: v1
kind: Secret
metadata:
  name: identity-key-checkout-api
  namespace: payments
type: Opaque
data:
  id: <base64-encoded-identity>
  secret-key: <base64-encoded-sk_ID>
  master-public-key: <base64-encoded-mpk>
\end{verbatim}

Kubernetes RBAC policies control which pods can access which Secrets. Secrets are encrypted at rest using the cluster's encryption provider \cite{k8sEncryptatRest} (e.g., using KMS integration with cloud provider key management services). We however, do re-assert that storing sensitive data in Kubernetes secrets can have certain issues such as increased key exposure and weaker leakage resilience compared to Secrets CSI driver-based solutions, and as such alternative solutions may be explored to secrets management.

\subsection{Service Discovery and Validation}

Service discovery in Kubernetes uses DNS and service objects. To integrate with identity-based TLS, we augment service discovery with identity metadata. When a client wishes to connect to a service, it queries the Kubernetes API or DNS to obtain the service name, then constructs the expected identity:
\[
ID_{\text{expected}} = \text{cluster} \| \text{namespace} \| \text{service} \| \text{current-epoch}
\]

The client derives $pk_{ID_{\text{expected}}}$ and uses it for encapsulation. Any entity possessing $sk_{ID_{\text{expected}}}$ can complete the handshake, which should only be pods belonging to the legitimate service.

To prevent identity misbinding attacks, the T-PKG can publish a signed identity registry, analogous to Certificate Transparency. This registry contains an append-only log of all issued identity keys with timestamps (stored in a Merkle-tree structure), enabling clients to verify that an identity was legitimately issued.

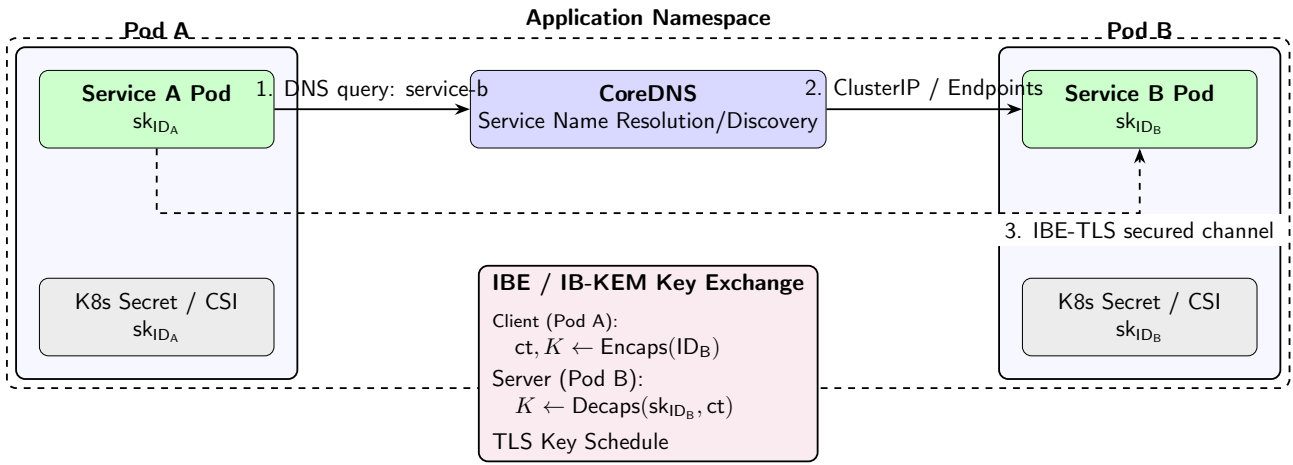
\begin{figure*}
    \resizebox{\textwidth}{!}{%
    \begin{tikzpicture}[
  font=\sffamily,
  node distance=2cm,
  box/.style={
    draw,
    rounded corners,
    align=center,
    minimum width=3.6cm,
    minimum height=1.2cm,
    fill=gray!5
  },
  pod/.style={
    draw,
    thick,
    rounded corners,
    inner sep=10pt,
    fill=blue!3
  },
  ns/.style={
    draw,
    dashed,
    thick,
    rounded corners,
    inner sep=14pt
  },
  arrow/.style={->, thick, >=Stealth}
]


\node[box, fill=green!20] (a-app) {
\textbf{Service A Pod}\\
$\mathsf{sk}_{\mathsf{ID_A}}$
};

\node[box, fill=gray!15, below=of a-app] (a-secret) {
K8s Secret / CSI\\
$\mathsf{sk}_{\mathsf{ID_A}}$
};


\node[box, fill=blue!15, right=3cm of a-app] (dns) {
\textbf{CoreDNS}\\
Service Name Resolution/Discovery
};


\node[box, fill=green!20, right=3cm of dns] (b-app) {
\textbf{Service B Pod}\\
$\mathsf{sk}_{\mathsf{ID_B}}$
};

\node[box, fill=gray!15, below=of b-app] (b-secret) {
K8s Secret / CSI\\
$\mathsf{sk}_{\mathsf{ID_B}}$
};


\node[draw,
      rounded corners,
      thick,
      fill=purple!8,
      align=left,
      minimum width=5.2cm,
      minimum height=2.6cm,
      below=1.8cm of dns] (kx) {

\textbf{IBE / IB-KEM Key Exchange}\\[4pt]
\footnotesize
Client (Pod A):\\
\quad $\mathsf{ct}, K \leftarrow \mathsf{Encaps}(\mathsf{ID_B})$\\[2pt]
Server (Pod B):\\
\quad $K \leftarrow \mathsf{Decaps}(\mathsf{sk}_{\mathsf{ID_B}}, \mathsf{ct})$\\[4pt]
\centering
TLS Key Schedule
};


\begin{scope}[on background layer]
  \node[pod, fit=(a-app)(a-secret),
        label=above:{\textbf{Pod A}}] {};
  \node[pod, fit=(b-app)(b-secret),
        label=above:{\textbf{Pod B}}] {};

  \node[ns, fit=(a-app)(a-secret)(dns)(b-app)(b-secret),
        label=above:{\textbf{Application Namespace}}] {};
\end{scope}


\draw[arrow] (a-app.east)
  -- node[above]{1. DNS query: service-b}
  (dns.west);

\draw[arrow] (dns.east)
  -- node[above]{2. ClusterIP / Endpoints}
  (b-app.west);

\draw[arrow, dashed]
  (a-app.south) -- ++(0,-1)
  -| node[below, fill=white]
  {3. IBE-TLS secured channel}
  (b-app.south);

\end{tikzpicture}
}
    \caption{Service Resolution and IBE-TLS communication}
    \label{fig:svc}
\end{figure*}

Figure \ref{fig:svc} shows Pod to Pod communication via IBE-TLS including Service Resolution using CoreDNS.

\subsection{Scalability Considerations}




The scalability of the proposed solution depends on the ability of the IBE-based PKI, realized through the Threshold Private Key Generator (T-PKG), and the resulting IBE-TLS handshake to scale with an increasing number of communicating parties and connection establishments. Since the T-PKG assumes a role analogous to that of a conventional PKI trust anchor, we expect comparable scaling characteristics, with potentially lower operational overhead due to the absence of certificate issuance, distribution, and validation. As the number of end entities requesting identity keys increases, the T-PKG must manage larger identity lists, support epoch-based key rotations, and ensure secure, distributed key storage to mitigate key escrow and leakage risks. Techniques such as batched key extraction—where multiple identity key requests are processed together—can help speed-up computation and mitigate scalability concerns.

The IBE-TLS handshake instantiated using ID-ML-KEM is computationally heavier than classical bilinear Diffie–Hellman–based IBE proposals (on the order of several times higher), but remains competitive with post-quantum TLS~1.3 in terms of bandwidth overhead. The computational cost of ID-ML-KEM operations is expected to be comparable to that of standard ML-DSA rather than ML-KEM, as practical implementations are likely to reuse the ML-DSA number-theoretic transform (NTT) stack. As a result, the IBE-TLS handshake exhibits similar computational characteristics and performance profiles to existing ML-DSA-based post-quantum deployments.
\section{Application to 5G Core Networks}

5G Core networks have increasingly started to use cloud-native deployments to cater to the extensive needs for ultra-reliable, low latency and edge deployments. The use of Network Function Virtualization (NFV) and  Software-defined networking (SDN) enables flexible, and programmable network, which can be dynamically instaniated. This approach is reinforced by the Service Based Architecture (SBA) of 5G Core Network (CN).

Consequently, the implementation of 5G network functions as microservices orchestrated by Kubernetes has become an attractive option, as evident by its usage in several open-source 5G cores, such as \textbf{free5GC}, \textbf{Open5GS}, etc. 

In an extension to this work, we describe how our protocol can be fit into the Post-Quantum 5G Core, \textbf{QORE}, described in \cite{QORE}. This core network builds upon free5GC and uses post-quantum TLS to secure the Service-Based Architecture. The system also deploys post-quantum PKI for certificate operations, all in a Kubernetes-based deployment.

\subsection{5G Service-Based Architecture overview}
5G Service-Based Architecture (SBA) provides a modular way to realize the 5G Core network, facilitating the cloud-native vision. This approach further enhances interoperability and compatibility, as the software systems are described on the basis of standardized service-based interfaces and protocol abstractions built over HTTP/2 and OpenAPIs.

Specifically, the Service-Based Architecture (SBA) defines communication between Network Functions (NFs) using a common service protocol based on OpenAPI, with messages encoded in JSON and transported over HTTP/2. Each NF exposes a set of standardized services, such as \texttt{Namf\_Communication} and \texttt{Nudm\_Sdm}, which are accessed through well-defined interfaces (e.g., \texttt{N11}, \texttt{N4}). An NF acts as a consumer when invoking services offered by another NF, and as a producer when providing services to peers. Access to these services is subject to authentication and authorization. The 3GPP specifications \cite{3gpp_33501} mandate the use of mutual TLS (mTLS), in conjunction with OAuth 2.0, to secure SBA communication. These modern security mechanisms align naturally with cloud-native environments, enabling integration with service meshes and identity providers commonly used in Kubernetes-based deployments.

Table \ref{tab:5g-nfs} lists the individual network function functionalities.
\begin{table*}[h]
\centering
\caption{Representative 5G Core Network Functions and Roles}
\label{tab:5g-nfs}
\begin{tabular}{ll}
\toprule
\textbf{Network Function} & \textbf{Primary Responsibility} \\
\midrule
AMF & UE registration, access and mobility management \\
SMF & PDU session management and UPF selection \\
UPF & User-plane packet forwarding and QoS enforcement \\
NRF & NF registration, service discovery, and authorization \\
UDM & Subscriber data management and credential storage \\
AUSF & UE authentication and key derivation support \\
PCF & Policy control for QoS, charging, and access rules \\
NSSF & Network slice selection for UE sessions \\
\bottomrule
\end{tabular}
\end{table*}

\subsection{Security Mechanisms in 5GC}
The 3GPP TS 33.501 provides as the security guidelines for 5G security, including authentication, confidentiality, integrity, and authorization
requirements for Service-Based Architecture (SBA) communication
\cite{3gpp_33501}. These requirements are realized through standardized
protocols and frameworks that operate at the application layer.

Mutual TLS~1.3, which is used to secure Service-Based Architecture (SBA) deployments, requires communicating parties to authenticate each other using certificates issued by a trusted authority. Instead of relying on pre-distributed peer certificates, this trust framework is typically established through a Public Key Infrastructure (PKI) that defines one or more trust anchors. These trust anchors, implemented as Certificate Authorities (CAs), are responsible for certificate issuance and lifecycle management, and enable peers to verify each other’s identities by validating certificate chains against the trusted CA set.

\subsubsection{How IBE can serve 5G SBA}
From the preceding discussion, it follows that while the semantics of Service-Based Architecture communication in the 5G Core remain unchanged under post-quantum security, the reliance on certificate-based authentication introduces increasing operational and performance overhead. IBE addresses this gap by providing certificate-free alternatives using identity-derived keys while preserving SBA and mutual authentication guarantees. With this, we begin with the application of IBE, beginning with the identity mapping for well-defined NFs, and then providing the PKI and communication overview.

Table \ref{tab:sba-tls} lists the major SBA APIs,  the participating Network functions and their TLS characteristics, implied by their operational roles.

\begin{table*}[h]
\centering
\caption{Representative SBA Interactions and TLS Impact in 5G Core}
\label{tab:sba-tls}
\begin{tabular}{lll}
\toprule
\textbf{SBA Interaction} & \textbf{Involved NFs} & \textbf{TLS Characteristics} \\
\midrule
NF Registration & All NFs (except NRF) $\leftrightarrow$ NRF & Short-lived, bursty \\
Service Discovery & SMF $\leftrightarrow$ NRF & Frequent, low payload \\
UE Registration & AMF $\leftrightarrow$ UDM & Latency-sensitive \\
Session Establishment & AMF $\leftrightarrow$ SMF & Multiple sequential calls \\
Policy Control & SMF $\leftrightarrow$ PCF & Repeated updates \\
Mobility Handling & AMF $\leftrightarrow$ AMF & Time-critical \\
\bottomrule
\end{tabular}
\end{table*}

\subsection{Identity Mapping for Network Functions}

Identity-based authentication aligns naturally with 5G Core deployments,
as Network Function (NF) identities are already explicitly defined,
centrally managed, and scoped to a single operator domain. In the 5G
Service-Based Architecture, each NF instance is uniquely identified and
registered with the Network Repository Function (NRF), making identity
an intrinsic system property rather than an external attribute.

We define NF identities as structured strings derived directly from
existing 3GPP identifiers:
\[
\texttt{ID} = \texttt{PLMN-ID} \| \texttt{NF-Type} \| \texttt{NF-Instance-ID} \| \texttt{Epoch}
\]

A short description of the fields used above is given in Table \ref{tab:5g-id-map},

\begin{table}[h]
\centering
\caption{Example Identity Mapping for 5G Network Functions}
\label{tab:5g-id-map}
\setlength{\tabcolsep}{4pt}
\begin{tabular}{ll}
\toprule
\textbf{Field} & \textbf{Description} \\
\midrule
PLMN-ID & Mobile network operator identifier \\
NF-Type & AMF, SMF, UPF, NRF, UDM, etc. \\
NF-Instance-ID & Unique instance or pod identifier \\
Epoch & Validity period / rotation window \\
\bottomrule
\end{tabular}
\end{table}

For example:
\begin{verbatim}
00101.AMF.amf-001.20250101
\end{verbatim}

This identity structure allows keys to be rotated by updating the epoch
component without requiring certificate revocation or distribution.
Importantly, all identities are scoped within a single operator domain,
making identity-based key issuance operationally appropriate. In total, the identity structure allows for:
\begin{itemize}
    \item Derivable from existing identifiers defined in 3GPP and registered with the Network Repository Function (NRF).
    \item Epoch field allows for frequent key rotation
    \item The identity string being human-readable serves well for debugging and auditing operations.
\end{itemize}

\subsection{CA replacement with T-PKG for 5G CN}
In the 5G Core, the Threshold Private Key Generator (T-PKG) directly replaces
the operator’s private Certificate Authority. Instead of issuing and
validating X.509 certificates for Network Functions, the T-PKG issues
identity private keys bound to 5G-defined NF identities.

The T-PKG is deployed within the operator domain and aligns with existing
5G trust assumptions, where all Network Functions are provisioned,
registered, and managed by a single mobile network operator.

\subsection{NF Registration and Identity Issuance}

NF registration in an IBE-based Service-Based Architecture follows the
standard 5G control-plane flow, with certificate-based authentication
replaced by identity-based authentication.

\paragraph{IBE-Based NF Registration Flow}
\begin{enumerate}
    \item The Network Function (NF) is provisioned with the T-PKG master
    public key (\texttt{mpk}) and the identity of the Network Repository
    Function (NRF).
    \item The NF establishes an IBE-TLS connection to the NRF using the
    NRF identity.
    \item The NF sends an \texttt{NFRegister} request containing its NF
    profile, without attaching a certificate.
    \item The NRF validates the request and forwards the NF identity
    string to the T-PKG.
    \item The T-PKG extracts the identity private key corresponding to
    the NF identity.
    \item The NF securely receives its identity private key over the
    authenticated channel.
    \item The NRF stores the NF profile together with the associated
    identity string.
\end{enumerate}

\subsection{Service Discovery and Peer Authentication}

After registration, Network Functions discover peers through the NRF and
authenticate using IBE-TLS.

\paragraph{Service Discovery and Authentication Flow}
\begin{enumerate}
    \item An NF queries the NRF for a required service (e.g.,
    \texttt{nudm-sdm}).
    \item The NRF returns matching NF profiles, including network
    location, supported services, and the NF identity string.
    \item The requesting NF derives the peer public key from the
    identity string and the T-PKG master public key.
    \item The NF establishes an IBE-TLS connection to the selected peer.
\end{enumerate}

A simplified view of the service discovery and authentication interaction
is shown below:
\begin{verbatim}
NF → NRF  : Query(service = nudm-sdm)
NRF → NF  : { endpoint, capabilities, identity = 
PLMN.NF-Type.NF-ID.Epoch }
NF → UDM  : IBE-TLS Handshake
(identity-derived public key)
\end{verbatim}

\subsection{Multi-PLMN and Roaming Considerations}

Cross-PLMN authentication between Network Functions can be supported
using federated T-PKG deployments.

\begin{itemize}
    \item \textbf{Bilateral master public key exchange:} Operators
    exchange T-PKG master public keys as part of roaming agreements,
    allowing public key derivation across PLMNs while retaining local
    trust control.
    \item \textbf{Hierarchical IBE (HIBE):} A roaming hub operates a root
    T-PKG, with operator-specific T-PKGs acting as subordinate
    authorities using hierarchical identity strings.
\end{itemize}

\subsection{IBE-TLS Compliance with 3GPP Security Requirements}

3GPP TS~33.501 specifies security requirements for 5G systems. Table
\ref{tab:3gpp-ibe-compliance} summarizes how IBE-TLS satisfies these
requirements using mechanisms equivalent to certificate-based TLS.

\begin{table*}[h]
\centering
\caption{IBE-TLS Compliance with 3GPP TS~33.501}
\label{tab:3gpp-ibe-compliance}
\begin{tabular}{lll}
\toprule
\textbf{Requirement} & \textbf{TS~33.501} & \textbf{IBE-TLS Mechanism} \\
\midrule
NF mutual authentication & 13.1 & Bidirectional IBE encapsulation \\
Confidentiality protection & 13.2.1 & TLS record layer unchanged \\
Integrity protection & 13.2.2 & TLS record layer unchanged \\
Replay protection & 13.2.3 & TLS handshake nonces \\
Authorization & 13.3 & Identity string carries scope \\
NF identity verification & 5.9.2 & Implicit via key possession \\
\bottomrule
\end{tabular}
\end{table*}

IBE-TLS satisfies 3GPP security requirements through equivalent security
mechanisms. While the authentication method differs from certificate-based
TLS, the resulting security properties and SBA semantics are preserved.

\subsection{Integration with QORE}

In QORE, post-quantum TLS is used to secure SBA interfaces, and a
post-quantum PKI is employed to issue and validate certificates for
Network Functions. Our proposed identity-based TLS mechanism can replace
certificate-based authentication within the same framework, while
retaining post-quantum confidentiality through ML-KEM-based key
exchange.

We reuse the framework described for 5G Core, in our integration of it with QORE. Briefly, QORE uses the following architecture:
\begin{itemize}
    \item Network Functions running as Kubernetes services/pods.
    \item Post-Quantum mTLS between the Network Functions.
    \item A private CA, running as a Kubernetes service, which issues
    post-quantum certificates to Network Functions.
\end{itemize}

Our integration replaces this architecture with the following:
\begin{itemize}
    \item Replacement of QORE’s private CA with a Threshold Private Key
    Generator (T-PKG).
    \item Modification of the Network Functions’ TLS stack to use
    IBE-TLS instead of certificate-based mTLS.
    \item Addition of NF-side identity key provisioning and lifecycle
    mechanisms, invoked during NF registration, for obtaining
    identity-derived private keys from the T-PKG (e.g., via an init
    container or sidecar). Re-registration for key rotation when the epoch is updated.
\end{itemize}

The integration preserves QORE’s security properties while eliminating
certificate issuance, distribution, and validation infrastructure.
Detailed implementation is planned as future work.

\subsection{Integration with Multi-Edge Deployments}
5G Core can be deployed at multiple edge locations, where each location can have its own T-PKG, and its own trust anchor, which ultimately lead to the network wide T-PKG, utilizing a Hierarchical IBE architecture for PKI.

\subsection{Benefits for 5G Core Deployments}

Applying identity-based TLS to post-quantum 5G Core networks yields
several advantages:
\begin{itemize}
    \item Reduced control-plane signaling overhead due to elimination of certificate transmission,
    \item Lower handshake latency for SBA APIs that are invoked frequently,
    \item Integrates seamlessly with post-quantum SBA
    \item The absence of conventional PKI-based systems, which often become complex in 5G deployments simplifies deployments and operational overhead.
\end{itemize}

All performance and scalability benefits discussed in earlier sections
directly apply to the 5G setting, where high connection churn and strict
latency budgets amplify the cost of certificate-based post-quantum TLS.

\section{Extension to Cloud-Native Infrastructure: Kubernetes}
This section describes a PKI architecture for the Kubernetes (K8s) control plane, demonstrating a complete alternative to conventional PKI systems based on digital signature schemes and X.509 certificates. This approach brings specific benefits of IBE, such as simplifying certificate management and enabling secure identification using user or device identities. It further enables the use of the IBE-TLS protocol between individual components, such as kube-apiserver, etcd, and kube-scheduler.

\subsection{Conventional PKI to IBE-based PKI}

Kubernetes PKI is structured using multiple \emph{logical} Certificate Authorities (CAs), each defining a distinct trust domain within the cluster. At minimum, these include a control-plane CA, an \texttt{etcd} CA, and a front-proxy CA used by the API aggregation layer~\cite{k8s_PKI}.

While these CAs are logically independent, they may be physically implemented using the same root key material (as is common in \texttt{kubeadm}-based deployments). Kubernetes nevertheless enforces separation through signer names, Extended Key Usage (EKU) constraints, and verification rules at the API server. Alternatively, a single root CA may delegate trust by issuing intermediate CAs for each domain.

\begin{table*}[t]
\centering
\caption{Kubernetes Component TLS Communication}
\label{tab:k8s_tls_components}
\renewcommand{\arraystretch}{1.2}
\begin{tabular}{|p{3.2cm}|p{3.2cm}|p{3.2cm}|p{2.2cm}|}
\hline
\textbf{Initiator} & \textbf{Responder} & \textbf{Certificate Used} & \textbf{mTLS} \\
\hline
kubectl / admin & API server & Client cert (control-plane CA) & Yes \\
\hline
Controller manager & API server & Client cert (control-plane CA) & Yes \\
\hline
Scheduler & API server & Client cert (control-plane CA) & Yes \\
\hline
kubelet & API server & Kubelet client cert & Yes \\
\hline
API server & kubelet & Kubelet serving cert & Yes \\
\hline
API server & etcd & etcd client cert & Yes \\
\hline
etcd peer & etcd peer & etcd peer cert & Yes \\
\hline
API server & aggregated API & Front-proxy client cert & Yes \\
\hline
\end{tabular}
\end{table*}

\begin{table*}[t]
\centering
\caption{Kubernetes Built-in Certificate Signers}
\label{tab:k8s_signers}
\renewcommand{\arraystretch}{1.25}
\begin{tabular}{|
  >{\raggedright}p{4.2cm}|
  p{3.2cm}|
  p{4.4cm}|
  p{3.2cm}|
}
\hline
\textbf{Signer Name} & \textbf{EKU} & \textbf{Permitted Subject / SAN} & \textbf{Auto-Approved} \\
\hline
kubernetes.io/kube-apiserver-client &
ClientAuth &
No fixed subject; subject restrictions enforced by admission plugins &
No \\
\hline
kubernetes.io/kube-apiserver-client-kubelet &
ClientAuth &
CN=system:node:\$NODE,
O=system:nodes &
Yes \\
\hline
kubernetes.io/kubelet-serving &
ServerAuth &
CN=system:node:\$NODE,
No SANs honored &
No \\
\hline
kubernetes.io/legacy-unknown &
Varies &
Unrestricted (legacy compatibility only) &
No \\
\hline
\end{tabular}
\end{table*}

\begin{table*}[t]
\centering
\caption{Kubernetes PKI Trust Domains}
\label{tab:k8s_pki_domains}
\renewcommand{\arraystretch}{1.2}
\begin{tabular}{|
  >{\RaggedRight}p{4cm}|
  p{6cm}|
}
\hline
\textbf{PKI Domain} & \textbf{Purpose} \\
\hline
Control-plane PKI &
API server authentication of users, nodes (kubelet), and controllers \\
\hline
etcd PKI &
Mutual authentication between etcd members and API server \\
\hline
Front-proxy PKI &
Delegated authentication for aggregated APIs \\
\hline
\end{tabular}
\end{table*}

\begin{figure}[t]
\begin{minipage}{0.8\textwidth}
\small
\begin{verbatim}
    apiVersion: certificates.k8s.io/v1
    kind: CertificateSigningRequest
    metadata:
      name: kubelet-client-node1
    spec:
      signerName: kubernetes.io/kube-api
      server-client-kubelet
      request: "<base64-encoded PKCS#10 CSR>"
      usages:
        - digitalSignature
        - keyEncipherment
        - clientAuth
\end{verbatim}
\end{minipage}
\caption{Kubernetes CertificateSigningRequest (CSR) resource specification}
\label{fig:csr_res}
\end{figure}

Upon initial startup, the kubelet authenticates to the API server using a bootstrap token, which grants permission to submit a CSR. The kubelet first requests a client certificate using the \texttt{kubernetes.io/kube-apiserver-client-kubelet} signer. Once approved, this certificate authenticates the kubelet as \texttt{system:node:\$NODE}. Kubernetes signers are entities which issue X.509 certificate material when CSRs are sent to them (Figure \ref{fig:csr_res}). The default K8s signers are listed in Table \ref{tab:k8s_signers}.

Optionally, the kubelet may request a serving certificate using the \texttt{kubernetes.io/kubelet-serving} signer. This certificate is presented to the API server when the API server initiates connections to the kubelet, enabling mutual TLS.

\subsubsection{Mutual TLS in Kubernetes}

All control-plane communication in Kubernetes is secured using mutual TLS. Both peers authenticate each other using X.509 certificates, and authorization decisions are derived from certificate subjects and organizations. There is no anonymous or one-way TLS within the control plane. Table \ref{tab:k8s_tls_components} provides an overview of the initiator and responder parties, the certificate used and issued by whom, and whether the connection is mTLS protected or not.











\subsubsection{IBE-based system for Kubernetes PKI}

We now describe an \textbf{IBE-based PKI system} for Kubernetes that preserves the existing security properties of the conventional Kubernetes PKI, including:

\begin{itemize}
\item Multi-domain trust distribution across control-plane components
\item A direct alternative to CertificateSigningRequests (CSRs)
\item Mutual authentication between servers and clients
\item An IBE issuer entity analogous to Kubernetes certificate signers
\end{itemize}

Our IBE-based PKI system enables IBE-TLS between Kubernetes components in a mutually authenticated manner, without relying on X.509 certificates or digital signature schemes. This reduces protocol and management overhead while providing equivalent authentication guarantees. In this design, certificates are eliminated and replaced by short identity strings, while private keys are derived by an identity issuer. 

The identity issuer serves as the IBE analogue of a conventional
Certificate Authority and is responsible for identity lifecycle
management, including issuance, rotation, and revocation. The IBE-based
PKI does not mandate a specific realization of this issuer. In the
simplest form, the issuer may be implemented as a single Private Key
Generator (PKG). In our setting, we instantiate the issuer as a
\emph{threshold} Private Key Generator (T-PKG), as described in previous
sections, to distribute trust and mitigate key escrow risks.

This is particularly relevant in the context of post-quantum cryptography, where the static, file-based PKI used by Kubernetes would otherwise grow substantially due to larger keys and signatures. In clusters with multiple trust domains, worker nodes, and clients, conventional PKI becomes increasingly difficult to manage securely. Our approach replaces certificates and signatures with identity-based key encapsulation, achieving implicit authentication through proof of private key possession.

\subsubsection{Multi-Domain Trust and Threshold-PKG}

We emphasize that thresholdization is not a fundamental requirement of
the IBE-based PKI design. The multi-domain trust structure described in
this section applies equally to both single-issuer PKG deployments and
threshold PKG (T-PKG) deployments. Thresholdization is introduced as an
optional enhancement to reduce single-point-of-compromise risk in
private cluster environments. Our later mechanisms --- Key provisioning/extraction, key lifecycle and trust, will work in the T-PKG domain, however, can be trivially adapted to a single PKG.

Now, we describe how the multi-domain trust distribution of Kubernetes can be reproduced using IBE. To mirror the existing Kubernetes PKI structure, we instantiate a minimum of three independent Threshold-PKGs (identity issuers), one for each logical trust domain:

\begin{itemize}
\item Kubernetes control-plane (API server domain)
\item etcd
\item Front-proxy (API aggregation layer)
\end{itemize}

Each T-PKG operates using a master public key (\texttt{mpk}) and a master secret key (\texttt{msk}), where the latter is distributed across multiple authorities according to a threshold policy. The \texttt{msk} is used to extract private keys for a given identity string. These identity issuers correspond directly to the logical CAs described in Table~\ref{tab:k8s_pki_domains}. Table \ref{tab:ibe_domains} lists the IBE issuers (T-PKGs) instantiated for Kubernetes PKI, which mimic the standard Kubernetes PKI. 
The secrets and the public key are stored at the following certificate paths:

\begin{figure}[H]
\begin{minipage}{0.60\textwidth}
\begin{verbatim}
/etc/kubernetes/pki/mpk.pem
/etc/kubernetes/pki/etcd/mpk.pem
/etc/kubernetes/pki/etcd/msk.pem
/etc/kubernetes/pki/apiserver-etcd-client.key
/etc/kubernetes/pki/apiserver.key
/etc/kubernetes/pki/apiserver-
kubelet-client.key
\end{verbatim}
\end{minipage}
\end{figure}
The above list is not exhaustive, but mimics the way K8s stores the PKI certificates and keys. The \texttt{apiserver-kubelet-client.key} is the key which the API server uses for communicating with kubelet as a client, this key is issued by the T-PKG at the API server. Similarly, for communicating with etcd as a client, the key is obtained from the etcd T-PKG, to be stored at \texttt{apiserver-etcd-client.key}. Thus, it is important to mention that same components exhibit different identities for different contexts. However, the difference is only concerned with authentication as a client or as a server, and the issuer. This is done to replicate the Extended Key Usage, and different certificates in standard PKI.

Additionally, the individual components may prefer to store their secret keys in one place, rather than in a threshold-manner. However, the secret key of the T-PKG's is kept unassembled and shared between the $n$ entities and thus, not shown here.

Table \ref{tab:ibe_identity_issuance} lists the Kubernetes components and the corresponding identities they are assigned.

\begin{table*}[h]
\centering
\caption{IBE Trust Domains Corresponding to Kubernetes PKI}
\label{tab:ibe_domains}
\renewcommand{\arraystretch}{1.2}
\begin{tabular}{|p{3.2cm}|p{4.2cm}|p{5.2cm}|}
\hline
\textbf{IBE Issuer (T-PKG)} &
\textbf{Kubernetes Equivalent} &
\textbf{Scope of Trust} \\
\hline
API Server T-PKG &
Kubernetes CA &
Authenticates API server clients and servers \\
\hline
etcd T-PKG &
etcd CA &
Secures etcd client and peer communication \\
\hline
Front-proxy T-PKG &
Front-proxy CA &
Delegated authentication for aggregated APIs \\
\hline
\end{tabular}
\end{table*}

\begin{table*}[h]
\centering
\caption{Identity Issuance in IBE-based Kubernetes PKI}
\label{tab:ibe_identity_issuance}
\renewcommand{\arraystretch}{1.2}
\begin{tabular}{|p{3cm}|p{4.5cm}|p{3.5cm}|p{3.5cm}|}
\hline
\textbf{Component} &
\textbf{Identity String} &
\textbf{Issuer (T-PKG)} &
\textbf{Usage} \\
\hline
API Server  &
\texttt{kube-apiserver} &
API Server T-PKG &
Server-side authentication \\
\hline
kubelet (node $i$) &
\texttt{kubelet:node-$i$} &
API Server T-PKG &
Client and serving authentication \\
\hline
Controller Manager &
\texttt{controller-manager} &
API Server T-PKG &
Client authentication \\
\hline
Scheduler &
\texttt{scheduler} &
API Server T-PKG &
Client authentication \\
\hline
etcd &
\texttt{etcd:node-$i$} &
etcd T-PKG &
Server and peer authentication \\
\hline
Front-proxy &
\texttt{front-proxy} &
Front-proxy T-PKG &
Delegated authentication \\
\hline
\end{tabular}
\end{table*}

\subsubsection{Identity Key Extraction and Bootstrap Integration}

We introduce a mechanism for identity key provisioning that serves as an
analogue to the Kubernetes \texttt{CertificateSigningRequest} (CSR) API.
For this purpose, we define a new resource, termed
\texttt{IdentityRequest}. This resource allows a component to request an
identity-bound private key from a designated issuer operating as a
Threshold Private Key Generator (T-PKG). 

\paragraph{Reusing Kubernetes TLS Bootstrapping with modifications.}
In standard Kubernetes TLS bootstrapping, a joining component (e.g., a
kubelet) is provisioned with a bootstrap token and the cluster CA
certificate. The CA certificate is used to authenticate the API server
and establish an initial TLS connection, over which the component submits
a CSR. Upon authorization and approval, the signed X.509 certificate is
returned and stored locally for subsequent mutual TLS connections.

Our design preserves the same control flow but replaces certificate-based
authentication with identity-based authentication using IBE-TLS, as
described earlier in this paper. Instead of relying on a CA certificate,
a joining component is provisioned out-of-band with the IBE master public
key (\texttt{mpk}) for the control-plane trust domain and the canonical
identity of the API server (e.g., \texttt{kube-apiserver}). The initial
secure channel is established by encapsulating directly to this
identity. Only the legitimate API server, possessing the corresponding
identity private key, can complete the handshake, providing implicit
server authentication without certificates.

Bootstrap credentials (e.g., bootstrap tokens or ServiceAccount tokens)
are then transmitted over this authenticated IBE-TLS channel and are
used exclusively for authorization, mirroring existing Kubernetes
bootstrapping semantics. The \texttt{system:bootstrappers} group is
critical in Kubernetes as it allows new nodes to authenticate and submit
registration requests to the API server; we reuse this group for IBE key
issuance, restricting members solely to creating
\texttt{IdentityRequest} objects.

\noindent
The transition from certificate-based bootstrapping to identity-based
bootstrapping follows a direct one-to-one mapping:
\begin{itemize}
  \item \texttt{system:bootstrappers} $\rightarrow$ Bootstrap identity (token / ServiceAccount)
  \item CSR creation $\rightarrow$ \texttt{IdentityRequest} creation
  \item CSR signer $\rightarrow$ IBE issuer
  \item CSR approval $\rightarrow$ \texttt{IdentityRequest} approval
  \item Node certificate $\rightarrow$ Node identity private key
  \item \texttt{system:node:\$name} $\rightarrow$ \texttt{kubelet:node-\$name}
\end{itemize}

\paragraph{Provisioning Flow.}
Identity key provisioning proceeds as follows:

\begin{enumerate}
  \item \textbf{Out-of-band bootstrap material:}  
  The joining component is provisioned with the control-plane
  \texttt{mpk}, the API server identity string, and a bootstrap token
  with restricted RBAC permissions. This is similar to the standard flow where API server's certificate or CA is already a part of the requesting party's trust store.

  \item \textbf{IBE-TLS channel establishment:}  
  The component establishes an IBE-TLS connection to the API server by
  encapsulating to the identity \texttt{kube-apiserver}. This replaces
  CA-based server authentication in TLS bootstrapping.

  \item \textbf{Identity request submission:}  
  Over the authenticated channel, the component submits an
  \texttt{IdentityRequest} using the Kubernetes-style REST API.

  \item \textbf{Authorization and approval:}  
  The API server authenticates the requester using the bootstrap
  credential and evaluates the request against RBAC policies and
  issuer-specific rules, analogous to Kubernetes CSR signers.

  \item \textbf{Threshold key extraction:}  
  Upon approval, the request is forwarded to the corresponding T-PKG.
  A threshold of T-PKG servers jointly participates in identity key
  extraction. No single server is capable of generating the identity
  private key on its own.

  \item \textbf{Secure key delivery:}  
  The requesting component establishes authenticated channels to the
  T-PKG servers, collects threshold responses, and locally reconstructs
  the identity private key.
\end{enumerate}

\begin{center}
\fbox{
\begin{minipage}{0.95\linewidth}
\small
\textbf{Example: Kubelet \texttt{node-01} Joining a Cluster}

\begin{itemize}
  \item \textbf{Identity:} \texttt{kubelet:node-01}
  \item \textbf{Issuer:} API Server T-PKG
  \item \textbf{Bootstrap credential:} \texttt{system:bootstrappers} token
\end{itemize}

\begin{enumerate}
  \item The kubelet is provisioned out-of-band with the control-plane
        master public key (\texttt{mpk}) and the API server identity.
  \item The kubelet establishes an IBE-TLS connection to
        \texttt{kube-apiserver} by encapsulating to its identity.
  \item Over this authenticated channel, the kubelet submits an
        \texttt{IdentityRequest} for \texttt{kubelet:node-01}.
  \item The API server authorizes the request using RBAC policies
        associated with the bootstrap identity.
  \item Upon approval, the API server invokes the T-PKG to extract the
        identity private key for \texttt{kubelet:node-01}.
  \item The kubelet receives the identity private key and subsequently
        authenticates to the API server using IBE-TLS.
\end{enumerate}
\end{minipage}
}
\end{center}

\paragraph{IdentityRequest API.}
The \texttt{IdentityRequest} resource closely replicates the structure of a
Kubernetes CSR, specifying the requested identity, intended usage, and
validity period:
\small
\begin{verbatim}
POST /apis/security.k8s.io/v1alpha1/identityrequests
Authorization: Bearer <bootstrap-token>
Content-Type: application/json

{
  "spec": {
    "issuer": "ibe.kubernetes.io/apiserver",
    "identity": "kubelet:node-01",
    "usage": ["client", "server"],
    "expirationSeconds": 86400
  }
}
\end{verbatim}
\normalsize
Authorization is enforced via standard Kubernetes RBAC policies, which
restrict which identities may be issued, their permitted usages, and
their maximum lifetime. 
As with TLS bootstrapping, approval may be
performed automatically by a controller or manually by an operator which does not affect the establishment/lifetime of the initial IBE-TLS/TLS connection.  For
example, similar to approving a CSR using

\begin{verbatim}
kubectl get csr
kubectl certificate approve <csr-name>
\end{verbatim}

\noindent{}an operator may inspect and approve identity requests using an analogous
workflow, such as

\begin{verbatim}
kubectl get identityrequests
kubectl identity approve <identityrequest-name>
\end{verbatim}

Each issuer enforces its own approval rules on identity names and usage,
in the same way Kubernetes certificate signers restrict CSR subjects
and key usages.

\paragraph{Key Delivery.}
Unlike certificate-based workflow, identity private keys are not
embedded in API object status fields. Instead, identity key material is
returned directly to the requesting component over authenticated
IBE-TLS channels established with the T-PKG/serving endpoint. A sample response
format is shown below:
\small
\begin{verbatim}
HTTP/2 200 OK
Content-Type: application/json

{
  "identity": "kubelet:node-01",
  "privateKey": "<base64-encoded id-private-key>",
  "mpk": "<base64-encoded master public parameters>",
  "expiration": "2026-01-01T00:00:00Z"
}
\end{verbatim}
\normalsize

The component stores the resulting identity private key in its local
secret store (e.g., node filesystem, in-memory keystore, or
CSI-based secret volume), according to deployment policy. The key is
then used directly within the IBE-TLS handshake for authentication.
\textbf{\\Note.} The T-PKG ensures that fewer than $t$ participating nodes
cannot reconstruct a client’s private key because of the underlying
threshold mechanisms. However, this quickly becomes a practical hassle:
in a fully thresholded setup, the client would need to contact each
T-PKG node, request its share, and then locally combine these shares to
reconstruct the correct identity private key.

To simplify this, we also consider a less secure but operationally
simpler alternative. In this mode, the API server/the serving party (or the PKG endpoint
itself) contacts the T-PKG nodes, collects the required threshold
shares, and reconstructs the private key ephemerally in memory. The
fully assembled key is then sent once to the client over an
authenticated IBE-TLS channel. The reconstructed key is never written
to disk and is immediately wiped from memory after delivery. \footnotetext{This mode is intended for environments where the API server or PKG endpoint is already a highly trusted component (e.g., control-plane nodes), and where reducing client complexity outweighs the additional trust placed in the issuer.}

\paragraph{Identity Lifecycle and Trust}
The time period for which identities remain valid, are handled through epochs, after which the identity must either be re-issued --- with a new private key, or get revoked. Issuers may refuse extraction for expired/revoked identity strings, and the T-PKG will maintain an append-only list of the issued identities, as highlighted previously.

This approach replaces the cluster CA certificate used in Kubernetes TLS
bootstrapping with the IBE master public key as the trust anchor for
initial authentication. The authenticity of the \texttt{mpk} and the
identity namespace is assumed to be established out-of-band, which is
appropriate for private, centrally administered environments. 

\paragraph{Extended Key Usage}
To account for extended key usage semantics in conventional PKI, components may request distinct identities for client-side and server-side authentication. For example, a kubelet may request identities of the form \texttt{kubelet:node-$i$/client} and \texttt{kubelet:node-$i$/server}. Identity scoping thus replaces X.509 EKU constraints.

In clusters with multiple worker nodes, kubelets obtain node-specific identities (e.g., \texttt{kubelet:node-01}). During initial startup, the kubelet authenticates using a bootstrap token, which authorizes it to submit an \texttt{IdentityRequest}. Once approved, the kubelet receives its identity private key and subsequently authenticates to the API server using IBE-TLS.

The notion of Kubernetes signers is preserved in the IBE-based system by mapping signer names to specific identity issuers. Each issuer enforces policy on permitted identities, usage types, expiration, and approval conditions, closely mirroring the semantics of Kubernetes CSR signers.

Table \ref{tab:auth_transition} provides a component-level mapping between the standard and our proposed alternative.

\begin{table*}[h]
\centering
\caption{Mapping from Conventional PKI to IBE-based PKI for Kubernetes Components}
\label{tab:auth_transition}
\renewcommand{\arraystretch}{1.2}
\begin{tabular}{|p{4cm}|p{5cm}|p{5cm}|}
\hline
\textbf{Component} & \textbf{Conventional PKI} & \textbf{IBE-based PKI} \\
\hline
kube-apiserver &
X.509 server cert; verifies client certs &
IBE server identity; implicit client authentication via IBE-TLS \\
\hline
kube-scheduler &
X.509 client cert to API server &
Identity key \texttt{scheduler} \\
\hline
kube-controller-manager &
X.509 client cert to API server &
Identity key \texttt{controller-manager} \\
\hline
etcd &
Server, peer, and client certificates &
Identity keys for all etcd communication \\
\hline
kubelet &
Client cert + serving cert &
Node-scoped identity keys \texttt{kubelet:node-$i$} \\
\hline
\end{tabular}
\end{table*}

\subsection{IBE-based PKI to IBE-TLS in Kubernetes Control Plane}

Kubernetes requires all connections between its components to be secured via TLS in order to maintain confidentiality and authenticity across the cluster. The usage of cluster-wide and multi-scoped CAs provides the basis for mutual authentication and zero-trust principles in K8s. Our description of an IBE-based PKI for Kubernetes follows the same design goals, enabling a mutually authenticated, certificate- and signature-free TLS protocol that is post-quantum in nature.

This protocol can be directly applied to control plane and node components, provided the system follows the IBE-based PKI operational model described in the previous section. In particular, each component holds an identity-based private key issued by the appropriate Threshold-PKG (API, etcd, or front-proxy domain), and each component is configured with the corresponding public system parameters (\texttt{mpk}) for the trust domain it communicates with.

We reuse our IBE-TLS construction and describe its instantiation for Kubernetes below, assuming a mutually authenticated connection between two components \(A\) and \(B\).

\begin{center}
\fbox{
\begin{minipage}{0.95\linewidth}
\textbf{IBE-TLS Handshake between Kubernetes Components}
\begin{enumerate}
  \item \textbf{Identity Selection:}
  Each component determines the expected peer identity string based on its role
  (e.g., \texttt{kube-apiserver}, \texttt{kubelet:node-01}, \texttt{etcd-peer-2}).

  \item \textbf{ClientHello + Identity Hint:}
  The initiating component sends a TLS ClientHello, including its identity string, ephemeral key share, and other protocol parameters/extensions. No certificate or signature is transmitted.

  \item \textbf{IBE Key Encapsulation:}
  Using the peer’s identity and the corresponding \texttt{mpk}, each party performs
  identity-based key encapsulation to derive a shared secret. Additionally, the server encapsulates the client's ephemeral key share, sharing the resulting ciphertext with the client.

  \item \textbf{Mutual Authentication by Key Possession:}
  Successful derivation of the handshake keys implicitly authenticates both parties,
  proving possession of the correct identity-based private keys.

  \item \textbf{Secure Channel Establishment:}
  Standard TLS record protection is established using keys derived from the IBE
  shared secret and the ephemeral key exchange secret.
\end{enumerate}
\end{minipage}
}
\end{center}

The security guarantees of the protocol remain unchanged. In particular, the protocol provides mutual authentication, forward secrecy (under the epoch-based identity construction), and confidentiality under post-quantum assumptions.

Table~\ref{tab:mtls-ibe} illustrates how mutual TLS inside a Kubernetes cluster is realized using IBE-TLS, and lists the identities used on each side of the connection. The identities are further subdivided into whether the peer (and the other peer) behaves as a client or as a server, for example, \texttt{kube-apiserver-client} and \texttt{kube-apiserver-server}.

\begin{table*}[t]
\centering
\caption{IBE-TLS Communication Between Kubernetes Components}
\label{tab:mtls-ibe}
\renewcommand{\arraystretch}{1.2}
\begin{tabular}{|l|l|l|l|}
\hline
\textbf{Initiator} & \textbf{Responder} & \textbf{Initiator Identity} & \textbf{Responder Identity} \\
\hline
kubectl & kube-apiserver &
\texttt{kubectl:<user>} &
\texttt{kube-apiserver} \\
\hline
kube-scheduler & kube-apiserver &
\texttt{kube-scheduler} &
\texttt{kube-apiserver} \\
\hline
kube-controller-manager & kube-apiserver &
\texttt{kube-controller-manager} &
\texttt{kube-apiserver} \\
\hline
kubelet (node-01) & kube-apiserver &
\texttt{kubelet:node-01} &
\texttt{kube-apiserver} \\
\hline
kube-apiserver & kubelet (node-01) &
\texttt{kube-apiserver-client} &
\texttt{kubelet:node-01} \\
\hline
kube-apiserver & etcd &
\texttt{kube-apiserver-client} &
\texttt{etcd-server} \\
\hline
etcd-peer-1 & etcd-peer-2 &
\texttt{etcd-peer-1} &
\texttt{etcd-peer-2} \\
\hline
kube-apiserver & front-proxy &
\texttt{kube-apiserver-client} &
\texttt{front-proxy} \\
\hline
\end{tabular}
\end{table*}

\section{Challenges and Limitations}

\subsection{Key Escrow and Trust}

IBE inherently introduces key escrow: the PKG can derive secret keys for any identity and thus decrypt all communications. While threshold PKG mitigates single-point compromise, the system still requires trust in the collective set of PKG servers. This is acceptable in private environments where the PKG is operated by the same entity deploying the services, but would be problematic for public internet applications. However, it is worth noting that due to the use of ephemeral key material
in the TLS handshake, \textbf{forward secrecy} is preserved. As a
result, compromise of long-term identity keys does not retroactively
expose past record-layer traffic, unless the communicating parties
themselves were compromised during the session.

The trust model shifts from "trust in certificate authorities not to mis-issue certificates" to "trust in PKG not to extract unauthorized keys." 

\subsection{Revocation Mechanisms}

Revocation in identity-based systems is more complex than in certificate-based PKI. Our epoch-based approach provides automatic expiration but lacks fine-grained, immediate revocation. Maintaining a revocation list at the T-PKG reintroduces some of the infrastructure overhead we sought to eliminate. 

Future work could explore more sophisticated revocation mechanisms such as accumulator-based approaches or integration with distributed ledgers for verifiable revocation status.

\subsection{Interoperability with Public PKI}

The identity-based system is designed for private environments and cannot directly interoperate with public internet services using standard certificate-based TLS. Hybrid deployments may require dual-stack support, where services maintain identity-based keys for internal communication and certificates for external communication. Hybrid IBE and conventional TLS protocols can also be realized to ensure dual support.

\subsection{Implementation and Integration to Kubernetes}

The absence of formally verified IBE-based PKI and IBE-TLS implementations, as
described in this paper, presents a considerable challenge for system
operators seeking to adopt the proposed design. Implementations for this shall first implement ID-ML-KEM, as a precursor to IBE-TLS and may realize the PKI with a PKG with a single node for simpler operations. 
Furthermore, applying
this architecture to the Kubernetes control plane requires non-trivial
architectural and networking changes, particularly around control-plane
authentication, bootstrap trust establishment, and integration with
existing API server and kubelet workflows. 

\subsection{Standardization concerns within 5G}
The flow presented in this paper for applying IBE to cloud-native 5G
systems deviates from current 3GPP-standardized authentication
mechanisms. As 5G deployments typically prioritize mature and
standardized solutions, such deviations may limit immediate adoption.
Nevertheless, the proposed approach serves as an experimental and
exploratory design point, and can inform hybrid, non-disruptive, and
interoperable deployments in which IBE-based authentication coexists
with classical certificate-based mechanisms.

\subsection{Performance of Threshold Operations}

Threshold trapdoor sampling introduces computational overhead compared to centralized key extraction. The distributed protocol requires multiple rounds of communication between the requesting entity and T-PKG servers. While this overhead occurs only during key provisioning (not during every TLS handshake), it could become a bottleneck in large-scale deployments with high pod churn rates.

\section{Future Work}

Ongoing advances in lattice-based cryptography and post-quantum protocol
design open several natural directions for extending this work, both at
the level of protocol assurance and cryptographic foundations, as well
as in practical deployment environments.

\paragraph{Formal verification.}

Formal verification of cryptographic protocols involves the use of
(often) domain-specific tools which, using machine-checked proofs,
analyze the security of cryptographic constructions, protocol flows,
and message formats. Such verification models are particularly effective
at identifying protocol-level flaws that may not be apparent through
informal reasoning or manual inspection, yet can have severe security
implications in practice.

Prior work has demonstrated the importance of formal analysis by
uncovering multiple vulnerabilities in earlier versions of the TLS
protocol, including the Lucky Thirteen attack~\cite{lucky_thirteen},
which exploited subtle timing leakage in TLS and DTLS record processing,
as well as logic-level failures in protocol design~\cite{bhargavan2014triple}
and imperfect forward secrecy guarantees in deployed TLS
implementations~\cite{adrian2015imperfect}. These results highlight the
necessity of rigorous formal verification for cryptographic and security
protocols.

In this work, we leave a full formal verification of the proposed
IBE-TLS handshake as future work. The protocol can be modeled in a
symbolic setting under the Dolev--Yao adversary model, where the attacker
controls the network but cannot break cryptographic primitives. Using
tools such as \texttt{hax} with a ProVerif backend, the identity-based
key encapsulation and TLS message flow can be expressed using equational
theories, enabling the verification of key security properties such as
mutual authentication, session key secrecy, forward secrecy, and
resistance to replay and unknown key-share attacks.

\paragraph{Cryptographic foundations and parameter refinement.}
Our construction builds on the ID-ML-KEM framework first proposed in ~\cite{id-ml-kem}, which has larger key and ciphertext sizes, and a larger modulus, leading to slightly higher decapsulation failures rates. This scheme is proven \texttt{IND-sID-CPA} secure in the Random Oracle Model under the decisional Ring-LWE assumption, which represents a weaker security notion than \texttt{IND-sID-CCA}.
The scheme also uses a discrete Gaussian sampler which uses an MNTRU \cite{CheonKimKimSon2019} trapdoor as the auxiliary information to extract a secret key corresponding to some ID. This would require a highly nontrivial implementation for resistance from Side-channel attacks. Additionally, while MNTRU provides improved
efficiency and tighter reductions compared to earlier designs, further
work is needed to systematically explore parameter choices, including
modulus selection, noise distributions, and trapdoor quality. The aim is to reduce the key and ciphertext sizes, keep the resulting scheme as close to ML-KEM as possible, and offer similar security guarantees and failure probabilities. 

\paragraph{Implementation}
An end-to-end implementation of the proposed architecture is planned to validate the design in practice. A reference implementation in Go (Golang) would realize the core protocol logic and supporting services, leveraging Go's concurrency model and native support for networking and cryptographic libraries. Such an implementation would enable evaluation of deployability, interoperability with existing TLS stacks, and performance in cloud-native environments.

Furthermore, extending this stack to QORE, shall help provide preliminary performance analysis, and present a reference architecture for the community.

\paragraph{Deployment and identity integration.}
From a systems perspective, integrating identity-based authentication
with existing service identity frameworks remains an important direction
for future work. Interfacing the proposed scheme with SPIFFE would allow
identity-based TLS authentication to align with standardized workload
identity models in cloud-native environments. Similarly, mapping
identity-based public keys to decentralized identity systems (DIDs)
could enable controlled cross-domain federation while preserving the
benefits of implicit authentication.

To mitigate practical risks associated with long-term key material,
future implementations should leverage hardware-backed security
mechanisms. Threshold-PKG secret shares and issued identity private keys
can be protected using Hardware Security Modules (HSMs) integrated into
Kubernetes-based deployments, preventing direct key extraction and
constraining cryptographic operations to trusted hardware. Existing
systems already demonstrate the feasibility of HSM-backed key
management in such environments~\cite{HSM_K8s}.

\section{Conclusion}

In this paper, we presented a certificate-free alternative for conventional
Private-PKI systems, which leverages Identity-Based Encryption based on
post-quantum, lattice-based primitives. We argued that Private PKI is, in
its current state, a mismatch for private environments where identities
are already strongly linked, centrally managed, and enforced by the
system. Motivated by the growing size and cost of post-quantum signature
schemes—which can substantially affect operational and network load—we
explored simpler alternatives that reduce PKI overhead using an
IBE-based PKI. This system enables parties to implicitly authenticate
themselves without exchanging certificates or digital signatures.

We demonstrated the applicability of this approach to latency-sensitive,
cloud-native systems, with particular emphasis on 5G Core networks. In
the context of the 5G Service-Based Architecture, we showed how
IBE-based PKI and IBE-TLS can replace certificate-based post-quantum
mTLS while preserving existing authentication semantics and security
requirements. Using the post-quantum 5G Core \textbf{QORE} as a concrete
example, we illustrated how identity-based authentication integrates
naturally with Kubernetes-based deployments, NRF-driven service
discovery, and operator-managed trust domains.

We further adapted the same architecture to the Kubernetes control plane,
presenting an alternative to the standard certificate-based PKI by
mapping conventional certificate roles to their IBE equivalents and
demonstrating identity issuance and revocation. We then extended this
framework to application workloads, where IBE-based PKI and IBE-TLS
operate directly between Pods and microservices, reusing the same trust
model and deployment assumptions.

While the approach introduces key escrow and requires trust in the PKG
infrastructure, these limitations are acceptable in private cloud
environments where the operator controls both the PKG and the services.
For such environments—including 5G core networks, microservice
platforms, and service meshes—the benefits of simplified key management
and reduced cryptographic overhead make identity-based TLS an attractive
alternative to certificate-based post-quantum TLS.

Future work includes formal verification, integration with emerging
identity standards, hardware-backed security enhancements, and
exploration of federated multi-cluster deployments. As post-quantum
cryptography transitions from research to deployment, alternative
approaches to certificate management will become increasingly important
for practical system design.

\bibliographystyle{IEEEtran}
\bibliography{references}

\end{document}